%% file: vamos_transactions.tex
\documentclass[journal]{IEEEtran}

\ifCLASSOPTIONonecolumn
\newcommand{\argmaxdist}{-12pt}
\else
\newcommand{\argmaxdist}{-3pt}
\fi

\usepackage[process=auto]{pstool}

\usepackage[utf8]{inputenc}

\usepackage[T1]{fontenc}

\usepackage{cite}

\usepackage{amsmath}

\usepackage{fixltx2e}

\usepackage{url}

\usepackage{microtype}

\usepackage{balance}

\usepackage{setspace}

\usepackage{xfrac}

\usepackage{booktabs}

\usepackage{dcolumn}

\usepackage{subfigure}

\usepackage{tikz}
\usetikzlibrary{arrows}
\usetikzlibrary{shadows}

\input{macros}

\newcommand{\subfigwidth}{0.48\linewidth}
\newcommand{\subfigspace}{0.0\linewidth}
\newcommand{\afterfigspace}{-2ex}
\ifCLASSOPTIONonecolumn
\newcommand{\figwidth}{0.55\linewidth}
\else
\newcommand{\figwidth}{\linewidth}
\fi

\ifCLASSOPTIONonecolumn
\else
\renewcommand{\baselinestretch}{0.98}
\fi

\EndPreamble
\begin{document}
%
\title{Receiver Concepts and Resource Allocation for OSC Downlink Transmission$^*$}

\author{\IEEEauthorblockN{Michael A.~Ruder$^1$, Raimund Meyer$^2$, Frank Obernosterer$^2$, Hans Kalveram$^2$, Robert Schober$^{1,2}$, and Wolfgang~H.~Gerstacker$^{1,2}$} \\
\IEEEauthorblockA{$^1$Institute for Digital Communications, Universit\"at Erlangen-N\"urnberg, \\ 
Cauerstraße 7, D-91058 Erlangen, Germany, \{ruder, schober, gersta\}@LNT.de\\
$^2$Com-Research GmbH, Wiesengrundstr. 4, D-90765 F{\"u}rth, Germany, \{raimund.meyer, frank.obernosterer, hans.kalveram\}@com-research.de}}
%
\maketitle

\begin{abstract}
Voice services over Adaptive Multi-user channels on One Slot (VAMOS)
has been standardized as an extension to the Global System for Mobile
Communications (GSM). The aim of VAMOS is to increase the capacity of GSM,
while maintaining backward compatibility with the legacy system.
To this end, the Orthogonal Sub-channels (OSC) concept is employed,
where two Gaussian minimum-shift keying (GMSK) signals are
transmitted in the same time slot and with the same carrier frequency.
To fully exploit the possible capacity gain of OSC,
new receiver concepts are necessary. In contrast to the base station,
where multiple antennas can be employed, the mobile station is typically equipped with only
one receive antenna. Therefore, the downlink receiver design is
a very challenging task. Different concepts for channel estimation,
user separation, and equalization at the receiver of an OSC downlink
transmission are introduced in this paper. Furthermore,
the system capacity must be improved by suitable downlink power and resource allocation algorithms.
Making realistic assumptions on the information available at the base station,
an algorithm for joint power and radio resource allocation is proposed. Simulation results show
the excellent performance of the proposed channel estimation algorithms, equalization schemes, 
and joint radio resource and power allocation algorithms in realistic VAMOS environments.

\end{abstract}

\unmarkedfootnote{$^*$ This paper has been presented in part at the IEEE International Symposium on Personal, Indoor and Mobile Radio Communications (PIMRC) 2009 and 2011, and the IEEE International Conference on Communications in China (ICCC) 2012.}

\IEEEpeerreviewmaketitle


\section{Introduction}
The Global System for Mobile Communications (GSM) is still by far the most popular cellular communication system worldwide.
Especially in emerging markets, there is the need for a major 
voice capacity enhancement of GSM in order to meet 
the demands of the customers. Different approaches to improve 
the spectral efficiency of GSM have been discussed.
For example, a tighter frequency reuse might be employed which, however, 
leads to increased interference from other users within the system.
Interference suppression techniques are required 
in order to avoid a performance degradation for small 
frequency reuse factors. To this end, single antenna 
interference cancellation (SAIC) algorithms have been developed,
e.g.~\cite{Gerstacker05,Chevalier2006}, exploiting the special 
properties of the Gaussian minimum-shift keying 
(GMSK) modulation used in GSM, which can be well approximated
by filtered binary phase-shift keying (BPSK). 
These algorithms are already employed
in commercial GSM terminals. Especially for downlink transmission they 
are highly beneficial because only a single receive antenna 
is required for interference suppression.

With the study item Multi-User Reusing One Slot (MUROS)
in 3GPP TSG GERAN an alternative to smaller reuse factors 
was proposed for voice capacity enhancement, cf.~\cite{nokia:07,chen:09}.
The MUROS study item led to the standardization of Voice services
over Adaptive Multi-user channels on One Slot (VAMOS) \cite{OSC_chap2011},
where the capacity is increased by deliberately overlaying two users in the same time slot and 
frequency resource within a cell. By this, in principle, the capacity can be doubled. 
Therefore, VAMOS is also one of the key enablers for the 
refarming of spectrum \cite{Paiva2012}. However, in the downlink,
only a single receive antenna can be assumed for both involved mobile stations (MSs), and for each of the 
MSs, the two overlaid transmit signals of the base station (BS) travel through the same propagation channel. 
With the aim of enabling a sufficiently good user separation and backward compatibility to the 
legacy GSM system, the Orthogonal Sub-channels (OSC) concept is
applied in the downlink of VAMOS \cite{TR45914:940}. 
In order to take into account that each user in a VAMOS user pair may
experience different propagation conditions (large-scale fading), 
different powers are assigned to both transmit signals 
resulting in a certain subchannel power imbalance ratio (SCPIR) \cite{nokia:07}.

Efficient receiver algorithms are necessary to cope 
with the interference in a VAMOS OSC downlink system.
In the downlink, joint estimation
of the SCPIR and the channel impulse response can be 
accomplished with only a small estimation performance 
loss compared to non-OSC GSM transmission \cite{Ruder2011}.
In the literature, few papers consider VAMOS downlink receiver algorithms.
In \cite{Vutukuri2011}, SAIC receiver algorithms for VAMOS downlink
transmission are proposed. However, these algorithms are only optimized to suppress
GMSK interferers. The algorithms proposed in this paper have an improved capability to also mitigate OSC interferers.

Since the interference situation is crucial for the overall user experience,
the performance of the receiver algorithms has to be evaluated in a network scenario.
Radio resource allocation (RRA) is considered here 
for the VAMOS downlink, which is more challenging than 
RRA for the uplink, where a standard multiple-input multiple-output (MIMO)
transmission scenario arises.
So far, only a very limited number of results on RRA for OSC and VAMOS are available.
In \cite{Molteni2011}, RRA for the VAMOS up- and downlink is studied. However, \cite{Molteni2011}
does not consider the problem of unknown interference caused by frequency hopping (FH) and
random speech activity.
It is assumed in \cite{Molteni2011} that the interference level
is known and the interference consists of only one dominant out-of-cell
GMSK interferer. In \cite{Ruder2012}, we have proposed
an RRA algorithm that takes into account unknown interference.
In the system model considered in \cite{Ruder2012}, the interference 
is caused by GMSK as well as OSC co-channel interferers.
Although the BS assigns users to VAMOS pairs and determines 
the frequencies and the transmit powers for the downlink, 
it is not possible to estimate the interference level for each user 
and in each burst due to FH.
The task of the proposed RRA algorithm is to minimize the required transmit power of the BS,
while trying to achieve some target frame error rate (FER) at each MS.
The algorithm in \cite{Ruder2012} is presented in more detail in this paper. Additionally
the influence of discontinuous transmission (DTX), where no
signal is transmitted during speech pauses, is analyzed, and
a hot spot scenario is investigated.

In summary, this paper makes the following major novel contributions:
\begin{itemize}
  \item The link and network aspects of downlink OSC transmission are presented for a common system model.
  \item A novel enhanced V-MIC receiver, requiring only one receive antenna, is derived. The superior FER performance compared
  to the receivers in \cite{Meyer2009} is confirmed by link level simulation results. 
  \item Various receiver algorithms are evaluated in network level simulations including RRA. The beneficial influence of advanced receiver algorithms 
  on the network capacity is studied, where the novel V-MIC exhibits significant capacity gains.
  \item Discontinuous transmission and a hot spot scenario, where only no-VAMOS interference is present, are analyzed
  in network level simulations. The capacity gain enabled by OSC downlink transmission in these scenarios is quantified by simulation results.
\end{itemize}

The paper is organized as follows. Section~\ref{sec:system_model} introduces
the system model of a VAMOS downlink transmission. In Section~\ref{sec:channelest}, 
channel estimation for the OSC downlink is discussed, and 
Section~\ref{sec:equalization} provides different equalization concepts 
and outlines the link-to-system 
mapping used for faster performance evaluation of the transmission system. 
A joint radio resource and power allocation algorithm is
presented in Section~\ref{sec:rra}. 
Sections~\ref{sec:num_res} and \ref{sec:conclusion} 
provide simulation results and conclusions, respectively.

\textit{Notation:}
$\expect\{\cdot\}$, $(\cdot)^T$, $(\cdot)^*$, and $(\cdot)^H$
denote expectation, 
transposition, conjugation, and
Hermitian transposition, respectively. 
$\mathrm{ln}(x)$
denotes the natural logarithm of $x$. 
Bold lower case
letters and bold upper case letters refer to column vectors 
and matrices, respectively.
$\mm{I}_X$ denotes the $X \times X$ identity 
matrix. 
${\rm Im}\{\cdot\}$ and ${\rm Re}\{\cdot\}$ 
stand for the imaginary and real parts of a complex number, respectively. 
$\left[\mm{A}\right]_{m,n}$ is
the element in the $m$th row and $n$th column of matrix $\mm{A}$.
$\mathcal{Z}^{-1}\{\cdot\}$ denotes the inverse $z$-transform. 
$< \cdot,\cdot>$ and $(\cdot) * (\cdot)$ stand for the inner product of two vectors
and the convolution operation, respectively.
$\lfloor x \rfloor$ denotes the largest integer not greater than $x$.
We use $f(x) = O(g(x))$ if and only if there exists a positive real 
number $C$ and a real number $x_0$ such that $|f(x)| \leq C \cdot |g(x)| \ \forall \ x > x_0$.


\section{System Model}
\label{sec:system_model}
In the considered model of an OSC downlink transmission, 
we focus on one specific cell. The BS of this cell serves a random 
number $N$ of users $i \in \mathcal{U} = \{1,\ 2,\ \ldots,\ N \}$.
Up to two user signals, corresponding to one pair, transmit in the same time slot
and the same frequency resource. According to the OSC concept \cite{nokia:07}, 
the first user of the pair (user $o \in \mathcal{U}$) 
and the second user (user $p \in \mathcal{U}$, $o \neq p$) have a phase difference of $90^\circ$.
In the equivalent complex baseband, 
the received signal at MS $o$  
after GMSK derotation can be written as
\ifCLASSOPTIONonecolumn
\begin{align}
	\tilde{r}_o[k] =& \sqrt{G_o} \sum_{\kappa=0}^{q_h} \tilde{h}_o[\kappa] \left(\sqrt{P_o}\, a_o[k- \kappa] + j \sqrt{P_p}\, a_p [k - \kappa] \right) 
	+ \tilde{n}_o[k] + \tilde{q}_o[k].
	\label{eq:received_sig}
\end{align}
\else
\begin{align}
	\tilde{r}_o[k] =& \sqrt{G_o} \sum_{\kappa=0}^{q_h} \tilde{h}_o[\kappa] \left(\sqrt{P_o} a_o[k- \kappa] + j \sqrt{P_p} a_p [k - \kappa] \right) 
		\nonumber \\ &
	+ \tilde{n}_o[k] + \tilde{q}_o[k].
	\label{eq:received_sig}
\end{align}
\fi
Here, the discrete-time channel impulse response (CIR) $\tilde{h}_o[k]$ of order $q_h$
is normalized to unit energy without loss of generality and
comprises the effects of GMSK modulation, the mobile channel from
the BS to the considered user $o$, receiver input filtering,
and GMSK derotation at the receiver. The path gain for transmission from the BS
to the receiver of user $o$ is denoted by $G_o$. It comprises the distance
attenuation and the large-scale fading, whereas $\tilde{h}_o[k]$ only characterizes the small-scale fading for user $o$.
We assume that $\tilde{h}_o[k]$ is constant for the duration of one burst (block fading model)
and unknown to the BS for RRA. $G_o$ changes only very slowly and is therefore 
assumed to be constant and known to the BS with sufficient accuracy 
by feedback from the MS. Interleaving is 
applied over one frame, which comprises $N_\mathrm{bursts}$ bursts.
$a_o[k]$ and $a_p[k]$ refer to the BPSK
transmit symbols of users $o$ and $p$, respectively,
and both symbols have variance $\sigma_a^2$. 
The average transmit powers for users $o$ and $p$ are denoted by $P_o$
and $P_p$, respectively.
$\tilde{n}_o[k]$ and $\tilde{q}_o[k]$ refer to discrete-time
additive white Gaussian noise (AWGN) of variance $\sigma_{\tilde{n}}^2$ and adjacent plus 
co-channel interference from other cells
at the receiver of user $o$, respectively.
The received signal $\tilde{r}_o [k]$ according to \eqref{eq:received_sig} is normalized 
by multiplication with $1/\sqrt{P_o}$ resulting in 
\begin{align}
	r_o[k] =& \sum_{\kappa=0}^{q_h} h_o[\kappa] \left(a_o[k- \kappa] + j\, b\, a_p [k - \kappa] \right) 
	+ n_o[k] + q_o[k].
	\label{eq:received_sig_norm}
\end{align}
Here, the overall CIR  of user $o$ is denoted by $h_o[\kappa] = \sqrt{G_o}\, \tilde{h}_o[\kappa]$, 
where the CIR of user $p$ is the CIR of user $o$ multiplied by $j$ and scaled by
a factor $b = \sqrt{P_p}/\sqrt{P_o}$, which is unknown at the receiver. 
$n_o[k]$ refers to AWGN with variance $\sigma_{n_o}^2$. 
The variance of the interference $q_o[k]$ is denoted by $\sigma_{q_o}^2$, 
assumed to be constant within each burst due to
synchronized network operation, and different 
for each MS. The received signal for user $p$
can be obtained analogously after exchanging $o$ and $p$ in 
\eqref{eq:received_sig_norm} and redefining $h_o[\kappa]$.
Single user GMSK transmission for user $o$
is included in \eqref{eq:received_sig} as a special case with $P_p = 0$ and
in \eqref{eq:received_sig_norm} with $b = 0$. 
The average signal-to-noise ratio (SNR) of user $o$ is given by 
\begin{align}
\mathrm{SNR}_o = (G_o P_o \sigma_a^2)/ \sigma_{\tilde{n}}^2.
\label{eq:snr}
\end{align}

Due to the fact that the power for each user within a pair
can be allocated individually, a power imbalance between 
the users arises. The corresponding SCPIR for 
user $o$ is defined as
\begin{align}
\mathrm{SCPIR}_o = 10 \log{}_{10}(P_o/P_p) = 10 \log{}_{10}(1/b^2).
\end{align}
$\mathrm{SCPIR}_o$ specifies the difference of both transmit
powers within one pair in $\mathrm{dB}$. Obviously,
$\mathrm{SCPIR}_p = -\mathrm{SCPIR}_o$ is valid. 
Due to receiver constraints, the power imbalance within one pair 
is limited to a maximum value \cite{TS45004:1100}. The maximal
allowed absolute value of SCPIR for RRA is denoted by 
$\mathrm{SCPIR}_\mathrm{max}$, i.e., 
$|\mathrm{SCPIR}_i| \leq \mathrm{SCPIR}_\mathrm{max}$ must be
valid for any user $i$. 

\section{Channel Estimation}
\label{sec:channelest}
As an initial task, we need to obtain estimates of the CIR and the SCPIR 
for the subsequent detection algorithms.
It should be taken into account that both user signals propagate to user $o$
through the same channel. We can rewrite \eqref{eq:received_sig_norm} 
in matrix-vector notation as 
\begin{eqnarray}
 \ma{r}_o	= \ma{A}_o \, \ma{h}_o + b \, \ma{A}_p \, \ma{h}_o + \ma{n}_o + \ma{q}_o,
\label{eq:transmission}
\end{eqnarray}
where $\ma{r}_o$ denotes the vector of the normalized received
symbols corresponding to the time-aligned training sequences
of both users, $\ma{A}_o$ and $\ma{A}_p$ represent $(N_\mathrm{tr}-q_h) \times (q_h+1)$ Toeplitz convolution matrices
corresponding to the training sequences of users $o$ and $p$, respectively, 
with training sequence length $N_\mathrm{tr}$, and $\ma{h}_o = \left[ \dsig{h_o}{0} \, \dsig{h_o}{1} \, \ldots \, \dsig{h_o}{q_h}  \right]^T$.
$\ma{n}_o$ and $\ma{q}_o$ are vectors containing the noise and interference contributions, 
respectively. 
For simplicity, factor $j$ in \eqref{eq:received_sig_norm} has been absorbed in $\ma{A}_p$. Furthermore, for 
channel estimation it is assumed that the composite impairment $\ma{w}_o = \ma{n}_o + \ma{q}_o$ is a Gaussian vector 
with statistically independent entries having variance $\sigma_{w_o}^2$.

\subsection{Joint ML Estimation of $\ma{h}_o$ and $b$}\label{sec:joint_ml_est}
The joint maximum-likelihood (ML) estimates for $\ma{h}_o$ and $b$ are obtained by minimizing the 
$L_2$-norm of the error vector $\ma{e} =  \ma{r}_o - \ma{A}_o \, \hat{\ma{h}}_o -
\hat{b} \, \ma{A}_p \, \hat{\ma{h}}_o$, where $\hat{\ma{h}}_o$ and $\hat{b}$ denote the estimated quantities. 
Differentiating $\ma{e}^H \, \ma{e}$ 
with respect to $\hat{\ma{h}}_o^*$ 
and $\hat{b}$ and setting the derivatives to zero results in the following two conditions for the ML estimates 
of $\ma{h}_o$ and $b$:
\begin{align}
\hat{\ma{h}}_o = \Big( 
{\big( \ma{A}_o^H + \hat{b}  \, \ma{A}_p^H \big)}
{\big( \ma{A}_o + \hat{b} \, \ma{A}_p \big)}
\Big)^{-1} 
{\big( \ma{A}_o^H     + \hat{b} \, \ma{A}_p^H \big)}
\, \ma{r}_o \label{mlchannelest}
\end{align}
%
and
\begin{align}
\hat{b}   = & \frac{1}{2} \left( \hat{\ma{h}}_o^H \, \ma{A}_p^H \, \ma{A}_p \, \hat{\ma{h}}_o 
\right)^{-1} \,  \left( \left( \hat{\ma{h}}_o^H \, \ma{A}_p^H \right) \, \left(\ma{r}_o - \ma{A}_o \, \hat{\ma{h}}_o
 \right) \right.  \nonumber\\
&  + \, \left. \left(\ma{r}_o^H -  \hat{\ma{h}}_o^H \, \ma{A}_o^H \right) \, \left( \ma{A}_p \, \hat{\ma{h}}_o 
\right)  \right).
\label{best}
\end{align}
Eqs.~\eqref{mlchannelest} and \eqref{best} may be also viewed as the ML channel estimate for
a given $b$ and the ML estimate of $b$ for a given channel vector, respectively \cite{crozier:91}. 
However, it does not seem possible
to obtain a closed-form solution for $\hat{\ma{h}}_o$ and $\hat{b}$ from the two coupled equations. Thus, a solution
might be calculated iteratively by inserting an initial choice for $\hat{b}$ in \eqref{mlchannelest}, using
the resulting channel vector for refining $\hat{b}$ via \eqref{best}, and so on, until convergence is reached.

\subsection{Blind Estimation of $\, b$}\label{sec:blind_chest}
In an alternative approach, $b$ is first estimated from the received vector according to
an ML criterion, assuming only knowledge of the channel statistics and both training sequences.  
Subsequently, the ML channel estimation is performed with the obtained $\hat{b}$ using \eqref{mlchannelest}.

Assuming $\ma{h}_o$ is a complex Gaussian vector with autocorrelation matrix 
$\ma{\Phi}_{h_o h_o} = \expect \{ \ma{h}_o \, \ma{h}_o^H \}$, 
the probability density function (pdf) of the received vector conditioned on $b$ may be expressed as
\begin{equation}
{\rm pdf} (\ma{r}_o \, | \, b ) =  \frac{1}{\pi^M \,  \det \left( \ma{\Phi}_{r_o r_o | b} \right)}
                         \exp \left( - \ma{r}_o^H  \, \ma{\Phi}_{r_o r_o | b}^{-1} \,  \ma{r}_o \right),
  \label{eq:pdf}
\end{equation}
where $M=N_\mathrm{tr}-q_h$ and $\ma{\Phi}_{r_o r_o | b} =  \expect \left\{ \ma{r}_o \, \ma{r}_o^H \, | \, b  \right\}$, 
\begin{equation}
\ma{\Phi}_{r_o r_o | b} =
\left(\ma{A}_o + b \, \ma{A}_p \right)  \, \ma{\Phi}_{h_o h_o} 
\left( \ma{A}_o + b \, \ma{A}_p \right)^H + \sigma_{w_o}^2 \, \ma{I}_{M}.
\label{eq:phi}
\end{equation}
The ML estimate for $b$ can be obtained by maximizing 
$\ln \left({\rm pdf} ( \ma{r}_o \, | \, b ) \right)$ using \eqref{eq:pdf} and \eqref{eq:phi}:
\ifCLASSOPTIONonecolumn
\begin{align}
\hat{b} =& 
\argmax{\argmaxdist}{\tilde{b}} \bigg\{ - \ma{r}_o^H \,  \ma{\Phi}_{r_o r_o \, | \, \tilde{b}}^{-1}\  \ma{r}_o 
- \ln \bigg[ \det \Big( \ma{\Phi}_{r_o r_o \, | \, \tilde{b}} \Big) \bigg] \bigg\} \nonumber\\
 =& \argmin{\argmaxdist}{\tilde{b}} \bigg\{ \ma{r}_o^H \, 
\bigg[ \bigg(\ma{A}_o + \tilde{b} \, \ma{A}_p \bigg) \,  \ma{\Phi}_{h_o h_o} 
\Big( \ma{A}_o + \tilde{b} \, \ma{A}_p \bigg)^H 	
+ \sigma_{w_o}^2 \, \ma{I}_{M} \bigg]^{-1} \, \ma{r}_o 
\nonumber \\ &
+ \, \ln \bigg[ \det \Big( \Big(\ma{A}_o + \tilde{b} \,  \ma{A}_p \Big) \, 
\ma{\Phi}_{h_o h_o} 
\Big( \ma{A}_o + \tilde{b} \, \ma{A}_p \Big)^H +  \sigma_{w_o}^2 \, \ma{I}_{M} \bigg) 
\bigg] \bigg\}.
\label{eq:fctmin}
\end{align}
\else
\begin{align}
\hat{b} =& 
\argmax{\argmaxdist}{\tilde{b}} \bigg\{ - \ma{r}_o^H \,  \ma{\Phi}_{r_o r_o \, | \, \tilde{b}}^{-1}\  \ma{r}_o 
- \ln \bigg[ \det \Big( \ma{\Phi}_{r_o r_o \, | \, \tilde{b}} \Big) \bigg] \bigg\} \nonumber\\
 =& \argmin{\argmaxdist}{\tilde{b}} \bigg\{ \ma{r}_o^H \, 
\bigg[ \bigg(\ma{A}_o + \tilde{b} \, \ma{A}_p \bigg) \,  \ma{\Phi}_{h_o h_o} 
\Big( \ma{A}_o + \tilde{b} \, \ma{A}_p \bigg)^H 	
\nonumber \\ &
+ \sigma_{w_o}^2 \, \ma{I}_{M} \bigg]^{-1} \, \ma{r}_o 
+ \, \ln \bigg[ \det \Big( \Big(\ma{A}_o + \tilde{b} \,  \ma{A}_p \Big) \, 
\ma{\Phi}_{h_o h_o} 
\nonumber \\ & \times 
\Big( \ma{A}_o + \tilde{b} \, \ma{A}_p \Big)^H +  \sigma_{w_o}^2 \, \ma{I}_{M} \bigg) 
\bigg] \bigg\}.
\label{eq:fctmin}
\end{align}
\fi

Minimization of the one-dimensional function in \eqref{eq:fctmin} might be performed
by a Golden section search technique \cite{brent:73}. 

It is interesting to compare the computational complexity of
the algorithms described in Section \ref{sec:joint_ml_est} and 
\ref{sec:blind_chest}. If we assume the joint ML estimation 
in Section~\ref{sec:joint_ml_est} needs $N_\mathrm{it}$ computations 
of \eqref{mlchannelest} and \eqref{best} to reach convergence, 
the computational complexity in terms of complex multiplications  
is dominated by the matrix multiplication in Eq.~\eqref{mlchannelest} 
and can be approximated as $O(N_\mathrm{it} \cdot (q_h+1)^2 (N_\mathrm{tr}-q_h) )$.
On the other hand, for the algorithm in Section~\ref{sec:blind_chest}, 
the dominant terms are the inversion and determinant operations. 
If we assume that also $N_\mathrm{it}$ iterations are necessary for the minimization of \eqref{eq:fctmin},
the computational complexity of the blind estimation algorithm can be approximated as 
$O(N_\mathrm{it} \cdot (N_\mathrm{tr}-q_h)^3)$.
Thus, we conclude that the computational complexity of the joint ML estimation
is lower than that of the blind estimation of $b$.
Simulations have shown that, in principle, both proposed estimation approaches for $b$ perform equally well 
under practical conditions. 

\section{Equalization and Interference Cancellation}\label{sec:equalization}
In the following, different equalization and interference cancellation algorithms
are introduced.
\subsection{Joint Maximum-Likelihood Sequence Estimation (MLSE)}
In noise limited scenarios, joint maximum-likelihood sequence estimation (MLSE) of sequences $\dsig{a_o}{\cdot}$ and $\dsig{a_p}{\cdot}$ 
(or a corresponding soft-output Viterbi algorithm 
\cite{fossorier:98} or Bahl--Cocke--Jelinek--Raviv (BCJR) algorithm \cite{koch:89} producing soft outputs) is optimum. For this, a Viterbi
algorithm (VA) in a trellis diagram with states
\begin{equation}
\dmatilde{S}{k} = [ \dsig{\tilde{a}_o}{k-1} \, \dsig{\tilde{a}_p}{k-1} \, \ldots \, 
 \dsig{\tilde{a}_o}{k-q_h} \, \dsig{\tilde{a}_p}{k-q_h}],
\end{equation}
where $\dsig{\tilde{a}_o}{\cdot}$, $\dsig{\tilde{a}_p}{\cdot}$ denote the trial symbols of the sequence estimator,
can be used. The branch metric for the state transitions is given by
\begin{equation}
\label{branchmetrics}
\dsig{\lambda}{k} = \bigg| \dsig{r_o}{k} - \sum\limits_{\kappa=0}^{q_h} \dsig{\hat{h}_o}{\kappa} \, 
\dsig{\tilde{a}_o}{k-\kappa} - j \, \hat{b} \, 
\sum\limits_{\kappa=0}^{q_h} \dsig{\hat{h}_o}{\kappa} \, \dsig{\tilde{a}_p}{k-\kappa} \bigg|^2.
\end{equation}
Equivalently, an MLSE for the modified 4QAM constellation $\{ -1- j \, \hat{b}, \, -1 + j \, \hat{b}, \, 
 +1- j \, \hat{b}, \, +1 + j \, \hat{b} \}$ can be applied. In both cases, the VA requires $4^{q_h}$ states.
%

\subsection{Mono Interference Cancellation (MIC)}
For reconstruction of the sequence of interest $a_o[\cdot]$, also a standard SAIC algorithm can be 
employed. Therefore, legacy MSs supporting Downlink Advanced Receiver 
Performance (DARP) phase I \cite{TS45005:9130} can be used also
for VAMOS without any change if legacy training sequences are employed. 
By a simple pure software update, also the eight new VAMOS training sequences \cite{nokia:07}
can be taken into account in a straightforward manner in an MS with SAIC receiver.
Thus, in the following, the MIC algorithm from \cite{saic_pat_pct, SAIC2005VTC, Gerstacker05} 
is briefly reviewed in the context of VAMOS. 

An arbitrary non-zero complex number $c$ is selected and a 
corresponding number $c^\perp={\rm Im}\{c\} - j \, {\rm Re}\{c\}$
is generated.
$c$ and $c^\perp$ may be interpreted as mutually orthogonal two-dimensional vectors.
The received signal is first filtered with a complex-valued filter with coefficients
$\dsig{f}{\kappa}$  
and then projected onto $c$, i.e., the real-valued signal 
\begin{equation}
\label{prooutput}
\dsig{y_o}{k} = {\cal P}_c \Big\{\sum\limits_{\kappa=0}^{q_f}
\dsig{f}{\kappa} \, \dsig{r_o}{k-\kappa} \Big\}
\end{equation}
is formed, where ${\cal P}_c\{x\}$ 
denotes the coefficient of projection of a complex number $x$ onto $c$, 
\begin{equation}
{\cal P}_c\{x\} = \frac{<x, \, c>}{|c|^2} = \frac{{{\rm Re}\{ x \, c^\ast \}}}{|c|^2}.
\end{equation}
It is shown in \cite{Gerstacker05}, that the filter
impulse response $\dsig{f}{\kappa}$ can be chosen for perfect elimination of 
signal contributions originating from $\dsig{a_p}{\cdot}$
(assuming $\dsig{a_o}{\cdot}$ is the desired sequence) if the filter order $q_f$ is sufficiently
high. After filtering and projection, $\dsig{a_o}{\cdot}$ can be reconstructed by trellis-based
equalization. An adaptive implementation of the MIC algorithm is also described in \cite{Gerstacker05} which
requires only knowledge of the training sequence of the desired user but no explicit channel
knowledge.

In typical urban (TU) environments, channel snapshots where a single tap dominates arise
frequently. Therefore, we consider the case $\dsig{h_o}{0} \neq 0$, $\dsig{h_o}{\kappa}=0, \, \kappa \neq 0$
($q_h=0$). The single effective channel tap $j \, b \, \dsig{h_o}{0}$ of the second user is rotated by 
90$^\circ$ compared to that of the first user. Therefore, in this case, orthogonal subchannels
result also at the receiver side. According to \cite{Gerstacker05}, suppression of the 
second user is possible without any loss in SNR, and 
${\rm SNR} =  2 \, |\dsig{h_o}{0}|^2 \, \sigma_a^2 / \sigma_{n_o}^2$ is valid after MIC if interference
from other cells is absent ($\dsig{q_o}{k}=0$). However, both subchannel contributions are not orthogonal
anymore at the receiver side for $q_h>0$,  and in general an SNR loss due to filtering and projection
cannot be avoided. Hence, as long as interference from other cells is absent, joint MLSE performs better 
than MIC which may be viewed as a suboptimum equalizer for QPSK-type signals in this case.

It should be noted that MIC is beneficial also for scenarios with several interferers \cite{Gerstacker05}. 
Here, the minimum mean-squared error (MMSE) filter found by adaptation is a kind of compromise
solution adjusted to the interference mixture. Given this and the fact that the interference created
by the other VAMOS pair user of the same BS is close to orthogonal to the desired user signal in many cases
for TU scenarios, it is expected that MIC performs better than joint MLSE in scenarios with
additional interference from other cells.

\subsection{MIC Receiver with Successive Interference Cancellation}
Because joint MLSE degrades significantly if external interference is present and DARP phase I receivers,
such as a receiver employing the MIC algorithm \cite{Gerstacker05}, 
typically exhibit a good performance only if the signal of the second VAMOS user 
is not much stronger than that of the considered user, more sophisticated schemes
are of interest for interference limited scenarios.
For this purpose, we exploit the fact that in contrast to the standard SAIC problem, the
training sequences corresponding to $\dsig{a_o}{k}$ {\em and} $\dsig{a_p}{k}$ are known at the MS,
and both signals are time aligned. Therefore, in principle, it is possible to reconstruct 
$\dsig{a_o}{\cdot}$ and $\dsig{a_p}{\cdot}$ in the same MS using two separate MIC algorithms. 

In a MIC receiver with successive interference cancellation (SIC), channel estimation 
according to Section~\ref{sec:channelest} is performed first. If $\hat{b}\ge b_0$ (e.g.~$b_0=1.0$), 
$\dsig{a_p}{\cdot}$ is reconstructed first
by combining the MIC algorithm with a subsequent trellis-based equalization yielding estimates $\dsig{\hat{a}_p}{\cdot}$.
In the next step, the contribution of $\dsig{a_p}{\cdot}$ is canceled from the received signal, resulting
in a signal
\begin{equation}
\dsig{r_{c,o}}{k} =  \dsig{r_o}{k} - j \, \hat{b} \, \sum\limits_{\kappa=0}^{q_h} \dsig{\hat{h}_o}{\kappa} \,
\dsig{\hat{a}_p}{k-\kappa},
\end{equation}
which is fed into another MIC and equalization stage in order to reconstruct $\dsig{a_o}{\cdot}$.
Because $\dsig{r_{c,o}}{k}$ contains no (or considerably reduced) contributions
from $\dsig{a_p}{\cdot}$, interference from other cells can be much better combated now by 
the second MIC. 

If $\hat{b}<b_0$, only a standard MIC is employed for reconstruction of $\dsig{a_o}{\cdot}$ because successive
interference cancellation most likely would suffer from error propagation.

In a typical implementation, the complexity of MIC with SIC is about 2.5 times higher than that of the standard
MIC, which is considered affordable in a typical modern MS.

\subsection{Enhanced VAMOS-MIC (V-MIC)}\label{sec:v-mic}

To further enhance the performance of the SIC receiver and to 
avoid the switching between different receiver types depending on $\hat{b}$, an
algorithm called VAMOS mono interference cancellation (V-MIC) can be used.
The performance of this scheme was first reported in \cite{comres2010}.
In the following, a detailed description of the algorithm is provided.
The basic idea of this enhanced receiver is to filter the received signal 
twice in parallel. 
In both filtering operations, only out-of-cell 
interference is suppressed, while the intersymbol interference
and interuser interference within the VAMOS pair are left in the signal.
Both prefiltered signals, representing the signals of users $o$ and $p$, are then 
fed to a joint MLSE. A similar idea has also been presented in \cite{Molteni2009}
for the uplink, but the algorithm in \cite{Molteni2009} relies on multiple receive antennas. The
V-MIC presented in the following requires only a single receive antenna, which is crucial
in the downlink.

\begin{figure}\begin{center}
\resizebox{\figwidth}{!}{
\input{equalizer_tikz}
}%
\caption{V-MIC structure for filter adaptation.}
\vspace*{\afterfigspace}
\label{abb:v-mic}
\end{center}\end{figure}
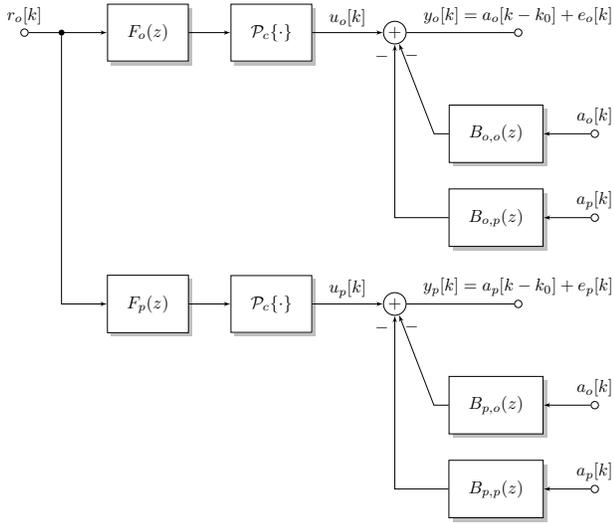
Fig.~\ref{abb:v-mic} shows the structure used for filter adaptation. 
The complex-valued received signal $r_o[k] = r_{o,\mathrm{I}} [k] + j\ r_{o,\mathrm{Q}} [k]$  is prefiltered
with two complex-valued filters $f_o[k] = f_{o,\mathrm{I}}[k]+j\ f_{o,\mathrm{Q}}[k]$
and $f_p[k] = f_{p,\mathrm{I}}[k]+j\ f_{p,\mathrm{Q}}[k]$. $F_o(z)$ and
$F_p(z)$ denote the $z$-transforms of $f_o[k]$ and $f_p[k]$, respectively. After
prefiltering, the resulting signals are projected onto $c$ and 
the real-valued signals $u_o[k]$ and $u_p[k]$ are obtained.
The prefilters are jointly optimized with the feedback filters
$b_{\nu,\mu}[k]$, where $\nu, \mu \in \{o,\, p\}$ and $B_{\nu,\mu}(z)$ denotes
the $z$-transform of $b_{\nu,\mu}[k]$. Filters $b_{o,o}[k]$ and $b_{p,p}[k]$
must be strictly causal, whereas $b_{o,p}[k]$ and $b_{p,o}[k]$ are 
causal filters\footnote{Strictly causal means that $b_{o,o}[0] = b_{p,p}[0] = 0$ in addition to causality. 
Therefore, the filter output only depends on past input values, 
but not on the current input value.}. 
The training sequences of users $o$ and $p$ in the VAMOS 
pair are both known at the receiver of user $o$
and used for adaptation of the respective filter.  


With these definitions and assumptions, the joint optimization of filters 
$f_o [k]$, $b_{o,o}[k]$, and $b_{o,p}[k]$ for the upper branch
in Fig.~\ref{abb:v-mic} can be achieved by minimizing
\ifCLASSOPTIONonecolumn
\begin{align}
\begin{split}
\sum_{k = 1}^{M} \Bigg| & \sum_{\kappa=0}^{q_f} f_{o,\mathrm{I}}[\kappa] r_{o,\mathrm{I}}[k-\kappa] 
-\sum_{\kappa = 0}^{q_f} f_{o,\mathrm{Q}}[\kappa] r_{o,\mathrm{Q}} [k- \kappa] - \sum_{\kappa=1}^{q_b} b_{o,o}[\kappa] a_o[k-k_0-\kappa] \\
&- \sum_{\kappa = 0}^{q_b} b_{o,p}[\kappa] a_p[k-k_0-\kappa] - a_o[k-k_0] 
\Bigg|^2
\end{split}
\label{eq:upper_branch}
\end{align}
\else
\begin{align}
\begin{split}
\sum_{k = 1}^{M} \Bigg| & \sum_{\kappa=0}^{q_f} f_{o,\mathrm{I}}[\kappa] r_{o,\mathrm{I}}[k-\kappa] 
-\sum_{\kappa = 0}^{q_f} f_{o,\mathrm{Q}}[\kappa] r_{o,\mathrm{Q}} [k- \kappa] \\ &- \sum_{\kappa=1}^{q_b} b_{o,o}[\kappa] a_o[k-k_0-\kappa] \\
&- \sum_{\kappa = 0}^{q_b} b_{o,p}[\kappa] a_p[k-k_0-\kappa] - a_o[k-k_0] 
\Bigg|^2
\end{split}
\label{eq:upper_branch}
\end{align}
\fi
with respect to the filter coefficients,
where $c=1$ has been assumed without any loss of generality
and $k_0$ is a decision delay which has to be optimized.
$q_f$ and $q_b$ are the orders
of the prefilter and feedback filter, respectively.
For the joint optimization of filters $f_p [k]$, $b_{p,o}[k]$, and $b_{p,p}[k]$ in the lower branch,
we need to minimize
\ifCLASSOPTIONonecolumn
\begin{align}
\begin{split}
\sum_{k = 1}^{M} \Bigg| & \sum_{\kappa=0}^{q_f} f_{p,\mathrm{I}}[\kappa] r_{o,\mathrm{I}}[k-\kappa] 
-\sum_{\kappa = 0}^{q_f} f_{p,\mathrm{Q}}[\kappa] r_{o,\mathrm{Q}} [k- \kappa] - \sum_{\kappa=0}^{q_b} b_{p,o}[\kappa] a_o[k-k_0-\kappa] \\
&- \sum_{\kappa = 1}^{q_b} b_{p,p}[\kappa] a_p[k-k_0-\kappa] - a_p[k-k_0] 
\Bigg|^2.
\end{split}
\label{eq:lower_branch}
\end{align}
\else
\begin{align}
\begin{split}
\sum_{k = 1}^{M} \Bigg| & \sum_{\kappa=0}^{q_f} f_{p,\mathrm{I}}[\kappa] r_{o,\mathrm{I}}[k-\kappa] 
-\sum_{\kappa = 0}^{q_f} f_{2,\mathrm{Q}}[\kappa] r_{o,\mathrm{Q}} [k- \kappa] \\ & - \sum_{\kappa=0}^{q_b} b_{p,o}[\kappa] a_o[k-k_0-\kappa] \\
&- \sum_{\kappa = 1}^{q_b} b_{p,p}[\kappa] a_p[k-k_0-\kappa] - a_p[k-k_0] 
\Bigg|^2.
\end{split}
\label{eq:lower_branch}
\end{align}
\fi
Minimizing \eqref{eq:upper_branch} and \eqref{eq:lower_branch}
is a standard least squares problem and finding the corresponding
optimal filter coefficients is straightforward.
After the filter adaptation, we have the following model
\begin{align}
\vv{u}[k] = \mm{B}[k] * \vv{a}[k-k_0] + \vv{e}[k],
\label{eq:sig_after_prefilter}
\end{align}
where $\vv{u}[k] = [u_o[k]\ u_p[k]]^T$, 
$\vv{a}[k] = [a_o[k]\ a_p [k]]^T$,
$\vv{e}[k] = [e_o[k]\ e_p[k]]^T$ (see Fig.~\ref{abb:v-mic} for the definition of $e_o[k]$ and $e_p[k]$) and 
$\mm{B}[k] = \mathcal{Z}^{-1} \{\mm{B}(z)\}$ with
\begin{align}
\mm{B}(z) = \left[
\begin{array}{cc}
1+B_{o,o}(z)& B_{o,p}(z)\\
B_{p,o}(z)& 1+B_{p,p}(z)
\end{array}
\right].
\end{align}

All entries of $\mm{B}(z)$ are causal. Thus, 
based on \eqref{eq:sig_after_prefilter},
a joint MIMO reduced state sequence estimation (RSSE) 
equalization can be performed \cite{zhang2005}.

\subsection{Link-to-System Mapping}
\label{sec:l2sysmap}
For FER performance evaluation of the receivers a link-to-system
mapping, similar to the one used in \cite{Brueck2004274}, is applied.
The idea of the link-to-system mapping is to approximate the
FER of a transmission with a mapping table, since it is too 
computationally complex to simulate each transmission
individually.
Our link-to-system mapping approach is based on two stages.
In the first stage, the raw
bit error rate is estimated for each burst comprising bits of a codeword representing a speech frame. 
This is done for each user, based on the power levels of all 
interferers (adjacent and co-channel), 
the power of the useful part of the received signal (including small-scale fading), and the SCPIR.
A five-dimensional look-up table is used for this which is also dependent
on the receiver algorithms since they differ
in their ability to separate the users of one pair.
For our simulations, we assumed that 
all MSs are VAMOS capable. One of the receivers 
described in {Sections~\ref{sec:equalization}-A-D} is employed. The look-up
tables with the raw bit error rates have been generated for different values of all
parameters by physical layer simulations of the respective receivers.

In the second stage, the FER is estimated based on the applied channel code and the 
mean value and the variance of the raw bit error rate of the bursts in the frame, cf. also \cite{Brueck2004274}.
Two-dimensional look-up tables were created for each GSM speech codec.
This stage is independent of the algorithm used for equalization and interference cancellation.
Combining these two stages efficiently models the interleaving and approximates the FER 
by obtaining the raw bit error rate for every burst from stage one and
using the mean and variance for stage two.

\section{Radio Resource Allocation}
\label{sec:rra}
For radio resource allocation (RRA), we assume that the BS
has knowledge of the large-scale fading
gain $G_i$ of each MS $i \in \mathcal{U}$ within its cell\footnote{This assumption
holds quite well in practice, since $G_i$ can be estimated accurately based on the  
received signal level (RxLev) measurements that are available at the BS.}. The small-scale fading and the 
interferer powers are unknown to the BS. For simplicity 
we assume that all MSs use the 
same speech codec\footnote{An extension to different speech codecs
is straightforward.}. 
In the following, 
RRA for the downlink case will be considered. 
The challenging parts of the RRA 
task for OSC transmission are the power allocation
for the pairs and the pairing of the users. 

The goal of our RRA optimization is, similar to \cite{Molteni2011},
the minimization of the required sum transmit power of the BS that serves one cell.
This is accomplished by finding the user pairing achieving this target.
In the considered cell in total $K$ logical channels are
available and $N$ users want to be served by the BS.
A logical channel, in contrast to a physical channel,
is not assigned to a specific frequency. 
There are two possibilities to use a logical channel,
either by employing conventional 
GMSK modulation, and thereby only transmitting one user signal,
or employing OSC modulation, where the logical
channel is ``split'' into two sub-channels for two users.
$\bar{K} \leq K$ channels are used for OSC transmission
and therefore $\bar{N} = 2 \bar{K}$ users are chosen to be paired.
The $\bar{N}$ paired users are collected in the set $\mathcal{N}$.
Section~\ref{sec:ups} gives details about different
strategies to determine the number of channels $\bar{K}$ used for OSC transmission. 
The set of all possible pairs composed of the $\bar{N}$ users of set $\mathcal{N}$ is denoted
by $\Pi$. There are $|\Pi| = \binom{\bar{N}}{2}$ possible pairs
in set $\Pi$. The goal of the optimization is to find
a pairing strategy 
$\mathcal{P} = \{ \mathcal{P}_1 ,\ \ldots, \ \mathcal{P}_{\bar{K}} \}$,
with $\mathcal{P}_k = \{o,\ p\}$, $\mathcal{P}_k \in \Pi$, and $k \in \{ 1,\ \ldots,\ \bar{K} \}$
that minimizes the total transmit power. The two users of the pair $\mathcal{P}_k$
on the $k$th logical channel are $o,\ p \in \{1,\ \ldots,\ \bar{N} \}$,
where $o \neq p$. The subsets must be disjoint, i.e.,
$\mathcal{P}_k \cap \mathcal{P}_{k'} = \emptyset$ for $k \neq k'$.

The optimization problem can be stated as
\begin{align}
\hat{\mathcal{P}} &=  \argmin{\argmaxdist}{\mathcal{P}} \sum_{i' \in \mathcal{N}} P_{i'}
\label{eq:opt_prob}
\end{align}
under the following constraints
for the users $\iota \in \{o,\ p\}$ in each pair $\mathcal{P}_k$ 
\begin{align}
\left\{
\begin{array}{l}
P(\mathcal{P}_k) = P_o + P_p \leq P_\mathrm{max} \\ 
|\mathrm{SCPIR}_\iota| \leq \mathrm{SCPIR}_\mathrm{max},\ \iota \in \{ o,\ p\} \\ 
\mathrm{FER}_\iota \leq \mathrm{FER}_\mathrm{thr},\ \iota \in \{ o,\ p\} 
\end{array} 
\right. .
\label{eq:constrains}
\end{align}
The first constraint limits the transmit power of 
each pair, $P(\mathcal{P}_k)$, to a maximum transmit power of $P_\mathrm{max}$.
The absolute value of $\mathrm{SCPIR}_\iota$ is limited to
$\mathrm{SCPIR}_\mathrm{max}$ for each user $\iota$ by the second constraint.
The last constraint demands a frame error rate below a threshold
of $\mathrm{FER}_\mathrm{thr}$.
There are two reasons why the total transmit power
is considered as a criterion. On the one hand, 
the interference to neighbor cells is 
decreased when the transmit power is lower. On the other hand, this also leads to a lower power consumption
of the BS, which will help to save energy in the network operation.
The optimum solution, given the assumed knowledge and the adopted criterion can be found with the  
algorithm proposed in Section~\ref{sec:paa}.
It is possible to construct some artificial scenarios, where a feasible
solution of the optimization problem cannot be found.
However, for the practical scenarios considered in Section~\ref{sec:num_res} 
with moderate requirements for the FER, it was always possible 
to find a feasible solution to the given problem.

The main challenge is the power allocation for each user.
To satisfy the $\mathrm{FER}$ constraint it is necessary to
allocate enough power to each user to guarantee some required signal-to-interference-plus-noise
ratio (SINR) at the receiver.
The interference power at the receiver cannot be estimated in RRA,
since the resource allocation is done independent of the other
cells that use the same frequencies. Furthermore, due to
frequency hopping the interference powers also change after each burst, whereas
the power allocation and the pairing are fixed for at least one frame,
which comprises up to $8$ bursts\footnote{The actual number of bursts in
one frame depends on the applied interleaving.}. Therefore, for RRA, it is proposed
to use a simplified mapping table, compared to 
that proposed in Section~\ref{sec:l2sysmap} for faster numerical evaluation.
This simplified RRA mapping table will be introduced in the following.

\subsection{Radio Resource Allocation Mapping Table}
Since the power level of the interferers is unknown to the
BS, some average interference power should be assumed for
RRA. Based on the mapping table 
described in Section~\ref{sec:l2sysmap} a simplified RRA
table is generated. As explained in Section~\ref{sec:l2sysmap},
the link-to-system mapping table needs as input the sum powers
of different interference types (adjacent channel interferers, 
co-channel GMSK and VAMOS interferers, etc.). The power of these 
different interference types is assumed to be $P_\mathrm{int}$ for each type.
Therefore, the RRA table for interference power $P_\mathrm{int}$ 
is only a function of $\mathrm{SCPIR}_{i'}$ and $\mathrm{SNR}_{i'}$
for each user $i' \in \mathcal{N}$,
\begin{align}
\mathrm{FER}_{i'} = f(\mathrm{SCPIR}_{i'}, \mathrm{SNR}_{i'}).
\label{eq:FER_user}
\end{align}

The table is generated for the specific number of
bursts used and some assumed power delay profile for the small-scale 
fading such as TU.

\subsection{Power Allocation Algorithm}
\label{sec:paa}
The necessary transmit power for a pair can be determined
by evaluating the RRA mapping table according to \eqref{eq:FER_user}
for different values of $\mathrm{SCPIR}_i$. 
The $\mathrm{SCPIR}_i$ values for the evaluation
are taken from the interval 
$[-\mathrm{SCPIR}_\mathrm{max}, \ \mathrm{SCPIR}_\mathrm{max}]$. 
For a given pair 
$\mathcal{P}_k = \{o,\ p\}$, the minimum $\mathrm{SNR}_\iota$ 
($\iota \in \{o,\ p\}$) necessary to fulfill the FER 
threshold $\mathrm{FER}_\mathrm{thr}$ for different
values of $\mathrm{SCPIR}_\iota$ can be interpolated from the RRA mapping table.
This is done by searching the smallest SNR value for the given
parameters that satisfies the FER threshold and the biggest SNR
value that does not satisfy the threshold. The necessary transmit power 
for the signal of user $\iota$, $P_\iota(\mathrm{SCPIR}_\iota)$ ($\iota \in \{o,\ p \}$), 
is then linearly interpolated from these entries of the mapping table
for the FER threshold value.
Since $\mathrm{SCPIR}_o = - \mathrm{SCPIR}_p$ and $G_o \neq G_p$, the power required
for each user within the pair $\mathcal{P}_k$ will be different.
The required transmit power of the VAMOS pair given the required 
transmit power for user $\iota$ of this pair and $\mathrm{SCPIR}_\iota$
can be calculated from
\begin{align}
	\tilde{P}_\iota(\mathrm{SCPIR}_\iota) = P_\iota(\mathrm{SCPIR}_\iota)/C(\mathrm{SCPIR}_\iota),
	\label{eq:power_pair}
\end{align}
where
\begin{align}
C(\mathrm{SCPIR}_\iota) = 10^{\mathrm{SCPIR}_\iota/10} / (1+10^{\mathrm{SCPIR}_\iota/10})
\end{align} 
is used to represent the individual contribution of user $\iota$ to the power of the pairing.
To satisfy the FER constraint for both users the total transmit power 
of pairing $\mathcal{P}_k$ is selected as the maximum
of the required transmit powers $\tilde{P}_\iota(\mathrm{SCPIR}_\iota)$ of both users
\begin{align}
	P(\mathcal{P}_k, \mathrm{SCPIR}_o) = \max(\tilde{P}_o(\mathrm{SCPIR}_o),\ \tilde{P}_p(-\mathrm{SCPIR}_o)).
	\label{eq:maxpower}
\end{align}

The SCPIR of user $o$ is chosen as
\begin{align}
\widehat{\mathrm{SCPIR}_o} = \argmin{\argmaxdist}{\mathrm{SCPIR}_o} P(\mathcal{P}_k, \mathrm{SCPIR}_o).
\label{eq:scpir_opt}
\end{align}
By that also the SCPIR chosen for user $p$, $\widehat{\mathrm{SCPIR}_p}$, is defined and the 
selected transmit power for pair $\mathcal{P}_k$ is
\begin{align}
	\hat{P}({\mathcal{P}_k}) = \min(P(\mathcal{P}_k, \widehat{\mathrm{SCPIR}_o}), P_\mathrm{max}).
\end{align}
The $\mathrm{min}(\cdot)$ operation ensures that
the maximum transmit power constraint is always fulfilled.
However, by limiting the transmit power to $P_\mathrm{max}$,
a $\mathrm{FER}$ higher than $\mathrm{FER}_\mathrm{thr}$ might result. 
This can be avoided by the selection of a codec with higher error
correction capability.

For all possible pairs in set $\Pi$, the lowest required transmit power 
and the corresponding SCPIR value are calculated. The required transmit
power and the respective SCPIR value are stored in matrices
$\mm{P}$ and $\mm{S}$ of dimension $\bar{N} \times \bar{N}$,
respectively. They are filled with $[\mm{P}]_{o,p} = [\mm{P}]_{p,o} = \hat{P}({\mathcal{P}_k})$ 
and $[\mm{S}]_{o,p} = \widehat{\mathrm{SCPIR}_o}$ and 
$[\mm{S}]_{p,o} = - \widehat{\mathrm{SCPIR}_o}$, respectively. 
Since our optimization problem is a weighted perfect matching 
problem in non-bipartite graphs \cite{papadimitriou1998combinatorial},
the \textit{Blossom Algorithm} can be applied on the 
transmit power matrix $\mm{P}$ to find
the pairing with the lowest sum power.
The algorithm finally returns the pairing $\hat{\mathcal{P}}$
that fulfills \eqref{eq:opt_prob} under the constraints in \eqref{eq:constrains}.
The values for the transmit powers of the users and the SCPIR can
be extracted from $\mm{P}$ and $\mm{S}$, respectively. 
The powers of users $o$ and $p$ of the $k$th pair 
can be computed as 
$P_o = [\mm{P}]_{o,p} \cdot C([\mm{S}]_{o,p})$ and $P_p = [\mm{P}]_{p,o} \cdot C([\mm{S}]_{p,o})$, respectively.

The power allocation for a single user with GMSK modulation is straightforward.
With an RRA mapping table for the FER that only depends on the SNR of the user, the
necessary transmit power can be easily assigned.

\subsection{User Pairing Strategies}
\label{sec:ups}
There are different possible strategies to decide which users should be paired
and which should transmit alone. The number of logical channels $\bar{K}$ that are used for OSC transmission
must be determined by some pairing strategy. As a
reference, the \textit{no-VAMOS} case is of interest where
$\bar{K} = 0$ logical channels use OSC transmission. Therefore, 
only up to $K$ randomly selected users can be served with GMSK modulation. 
If $N > K$, the remaining users will be blocked and cannot be served.
To see the effect of \textit{pure VAMOS}, where as many users as possible are paired,
$\bar{K} = \min(K, \lfloor N/2 \rfloor)$ OSC transmissions are used.
If $N<2K$ and $N$ odd, one randomly selected user
transmits without VAMOS. If $N>2K$, $N-2K$ users are blocked. For the case of $N<2K$
some channels may remain unused, since there are more logical channels than pairs.

Another option is to only pair users
if $N > K$. The \textit{pair only if otherwise blocked} (POOB) strategy will 
be identical to the no-VAMOS case if $N \leq K$ and is identical to pure VAMOS if
$N \geq 2K$. For $K < N < 2K$, $\bar{K} = N-K $ channels are 
used for OSC transmission and $K - \bar{K}$ channels are used with GMSK modulation.
For all strategies, the users that are served with OSC or GMSK transmission 
are chosen at random.

To reduce the complexity of the RRA, it is also possible to use a
random pairing instead of optimizing the pairing according to \eqref{eq:opt_prob} and \eqref{eq:constrains},
for which $|\Pi|$ possible pairs
have to be evaluated. 
For the optimization
of SCPIR, one can set $\mathrm{SCPIR}_i = 0$~dB $\forall i \in \{1,\ \ldots,\ N\}$, 
which corresponds to equal user powers within one pair. 
Another possibility is 
to combine random pairing with SCPIR optimization for each of the randomly formed pairs.
The computational complexity is much lower than for optimum user pairing, since only for $\bar{K}$
pairs the optimum SCPIR must be determined according to \eqref{eq:scpir_opt}.


\section{Simulation Results}
\label{sec:num_res}
This section presents simulation results for channel estimation, receiver
algorithms and network simulations.
\subsection{Channel Estimation and Receiver Algorithms}
\label{sec:num_res_rec_alg}
\begin{figure}\begin{center}
\resizebox{\figwidth}{!}{%
\psfragfig{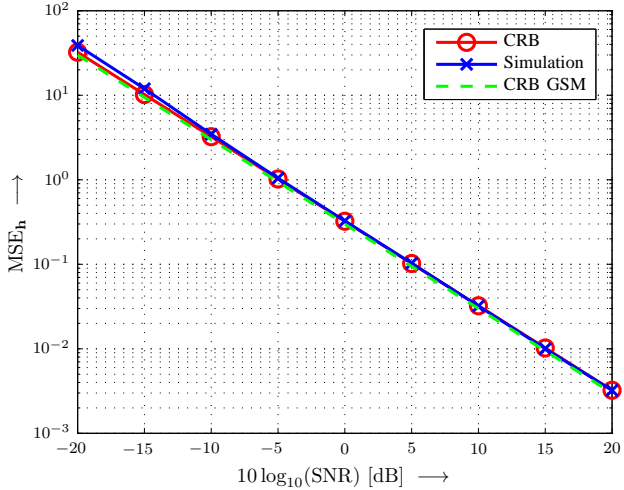}}
\caption{Sum MSE versus SNR for the estimation of the
 channel coefficients ($b = 1$).}%
\label{abb:crb_h}%
\vspace*{\afterfigspace}
\end{center}\end{figure}%
In the following, simulation results for channel estimation and an evaluation of 
the performance of different receiver algorithms are presented. %
In Fig.~\ref{abb:crb_h}, the sum mean-squared error (MSE)\footnote{
The sum MSE is defined as the sum of the MSEs of the estimates
of the individual channel taps.} 
for VAMOS channel estimation is compared
with the Cramer-Rao Lower Bound (CRB) for conventional GMSK 
transmission \cite{DeCarvalho1997} and the CRB for VAMOS, derived in \cite{Ruder2011}.
The joint ML estimation from Section~\ref{sec:joint_ml_est} is applied for 
VAMOS channel estimation. The SCPIR value is set to $b=1$,
while $5000$ different channel 
impulse responses of order $q_h = 5$ have been generated with 
independent taps drawn from a random complex
normal distribution with zero mean and variance $\sigma_h^2 = 1/(q_h+1)$.
The $\mathrm{SNR}$ of user $o$ relevant for channel estimation for 
OSC transmission is $\mathrm{SNR}^{\mathrm{OSC}} ={(1+b^2) \sigma_a^2}/{\sigma_{n_o}^2}$,
while for conventional GSM transmission the relevant $\mathrm{SNR}$
is $\mathrm{SNR}^{\mathrm{GMSK}} = { \sigma_a^2}/{\sigma_{n_o}^2}$,
to ensure a fair comparison. 
For OSC transmission, we can exploit the power of both subchannels since 
both training sequences are known. However, for non-OSC transmission,  
only the transmit power of the user of interest can be used since all 
interferers have unknown training sequences. 
Training sequence code (TSC) $0$ and the corresponding VAMOS TSC $0$ \cite{TR45914:940} have
been used for the simulations.

The proposed estimator matches the CRB closely for a broad range of 
SNR values. Only for low SNRs a minor degradation is visible. A loss in
channel estimation accuracy compared to channel estimation for conventional
GSM transmission is barely visible for the considered SCPIR value of $0$~dB.
In \cite{Ruder2011} also results for the estimation of the SCPIR can be found which
show a similarly good MSE performance.


%
%

\begin{table}%
\centering
\caption{MTS-1 and MTS-2 interference scenarios \cite{TR45914:940}.}
\begin{tabular}{llD{,}{,}{13}}
\toprule  
         Scenario & Interfering signal & \multicolumn{1}{l}{Interferer relative power}\\ 
\midrule   MTS-1    & Co-channel 1 & 0~\text{dB} \\
\midrule   MTS-2    & Co-channel 1 & 0~\text{dB}\\
										& Co-channel 2 & -10~\text{dB}\\
										& Adjacent channel 1   & 3~\text{dB}\\
										& AWGN         & -17~\text{dB}\\
\bottomrule 
\end{tabular}
\label{tab:mts} 
\end{table}%
For evaluation of the different receiver algorithms, a TU channel profile is
considered for an MS speed of $3$~km/h (TU3). Ideal frequency hopping
over $N_\mathrm{bursts} = 8$ bursts is used in the $900$~MHz band. 
Again, TSC $0$ and VAMOS TSC $0$ have been used.
Speech transmission with adaptive multirate (AMR) speech coding with full
rate (TCH/AFS 5.9 codec) is investigated. 
It is shown in \cite{Paiva2010} that mean opinion score (MOS) gains 
for speech quality can be achieved by using full rate speech coding 
in conjunction with OSC instead of half rate speech coding with non-OSC transmission.
For the interference from other cells, the MTS-1 and MTS-2 models from \cite{TR45914:940}
have been used. In MTS-1, only a single co-channel interferer is present, 
while MTS-2 defines an interference mixture. The details are given in Table~\ref{tab:mts}.
All interferers use GMSK modulation, and their TSCs are randomly chosen from
the eight specified GMSK TSCs. The interferers are synchronized with the desired signal.

In the receiver, channel estimation
and filter adaptation were used and a time slot based frequency offset
compensation was active. Receiver impairments such as phase noise
and I/Q imbalance were taken into account, and typical values for an
implementation were selected, cf. \cite{comres2010}.

\begin{figure}\begin{center}%
\resizebox{\figwidth}{!}{%
\psfragfig{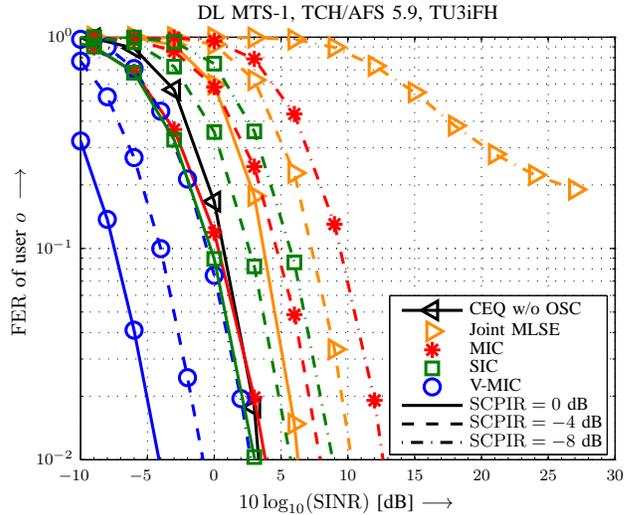}}
\caption{FER of user $o$ versus $\mathrm{SINR}$ for MTS-1 scenario and different receivers.}%
\label{abb:MTS1_AFS59}
\vspace*{\afterfigspace}
\end{center}\end{figure}%
In Fig.~\ref{abb:MTS1_AFS59}, the FER of user $o$ after channel
decoding versus $\mathrm{SINR}$ 
is shown for joint MLSE, MIC, SIC, and V-MIC for the MTS-1
scenario. In general, $\mathrm{SINR}$ denotes the power of the
OSC signal of both users, received by MS $o$, divided by
the power of co-channel interferer 1 according to Table~\ref{tab:mts}.
Results for different SCPIR values are shown. Also depicted
is the performance of the conventional GSM equalizer (CEQ) without 
interference suppression capabilities for
a pure GMSK transmission. 
For an SCPIR value of $0$~dB 
MIC and SIC perform very similar, but the V-MIC shows a significant
improvement of about $8$~dB in SINR for the same FER. A very
similar gain of V-MIC compared to MIC and SIC can also be observed for lower SCPIR values.
One can conclude from Fig.~\ref{abb:MTS1_AFS59} that it is possible to cancel one interferer quite well 
with the advanced V-MIC structure, while the degradation of the other
receivers is more significant for low SCPIR values.
The joint MLSE receiver is beneficial for noise limited scenarios, cf. \cite{Meyer2009},
but for interference limited scenarios this receiver has a very poor FER performance.
This can be explained by the fact that the joint MLSE cannot mitigate 
interference but treats it as noise. Therefore, a severe performance degradation
occurs in particular for low SCPIR values. 

\begin{figure}\begin{center}%
\resizebox{\figwidth}{!}{%
\psfragfig{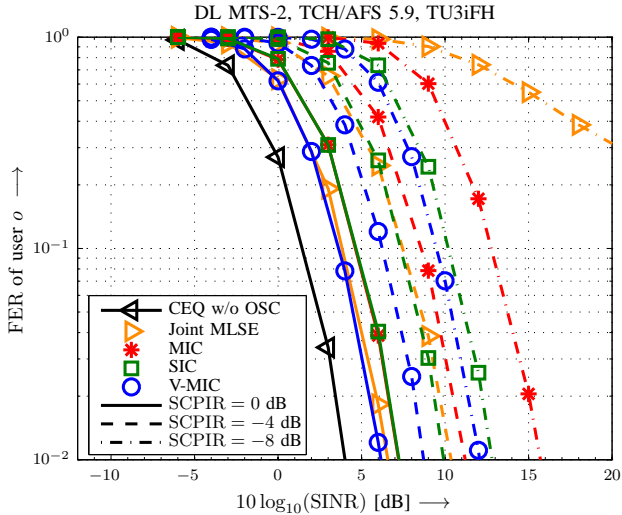}}
\caption{FER of user $o$ versus $\mathrm{SINR}$ for MTS-2 scenario and different receivers.}%
\label{abb:MTS2_AFS59}
\vspace*{\afterfigspace}
\end{center}\end{figure}%
For the MTS-2 scenario considered in Fig.~\ref{abb:MTS2_AFS59}, not only one
GMSK interferer is present but multiple interferers, cf. Table~\ref{tab:mts}.
The novel V-MIC achieves a gain of $1$~dB compared to the SIC and MIC receivers 
for an SCPIR of $0$~dB and the FER is also better than that for joint MLSE. 
The lower gain of V-MIC compared to the MTS-1 scenario 
is due to the higher number of interferers. Such an interference mixture cannot be combated 
as well as a single interferer. The loss of MIC compared to SIC and V-MIC, 
respectively, increases for lower SCPIR values, while the gain
of V-MIC is still $1$~dB compared to SIC. The performance of joint MLSE degrades severely for 
lower SCPIR values which has been also observed for MTS-1.

\subsection{Network Simulations}
\label{sec:network_sims}
\begin{table}[ht]
\centering
\caption{Simulation parameters.}
\begin{tabular}{|p{0.54\linewidth}|p{0.33\linewidth}|}
\hline 
 cell radius [m]& $500$\\ \hline 
 sectors per cell & $1$\\ \hline
 reuse factor & $12$\\ \hline
 number of clusters & $9$\\ \hline
 small-scale fading & Typical Urban (TU)\\ \hline
 pathloss model & UMTS 30.03, Vehicular Test Env.\\ \hline
 distance attenuation coefficient & $3.76$\\ \hline
 gain at $1$ m distance [dB] & $-8.06$\\ \hline
 standard deviation for the log-normal fading [dB] & $8$\\ \hline
 channels available in each cell & $K = 8$\\  \hline
 max. transmit power [dBm] & $P_\mathrm{max} = 30$\\ \hline
 noise power [dBm] (thermal noise + noise figure) & $-119.65 + 8$\\ \hline
 $\mathrm{SCPIR}_\mathrm{max}$ [dB] & $12$\\ \hline
 number of bursts for frequency hopping & $N_\mathrm{bursts} = 4$\\ \hline
 power of each interference type relative to noise power [dB] & $P_\mathrm{int} = \{10,\, 13,\, 15\}$ \\ \hline
 speech codec & AHS 5.9 (half rate)\\ \hline
 carrier frequency [MHz] & 900 \\ \hline
\end{tabular}
\label{tab:sim_par}
\end{table}

Table~\ref{tab:sim_par} gives an 
overview of the parameters used for the simulation results for RRA.
Only one of the $8$ periodic GSM time slots has been
simulated. Ideal random frequency hopping over all
available frequencies in each cell is used. In contrast to
Section~\ref{sec:num_res_rec_alg}, here the AMR half
rate speech codec with a bit rate of $5.9$~kbps (AHS 5.9) is used for all simulations
to maximize the network capacity.
All cells are frame synchronized and $\mathrm{FER}_\mathrm{thr} = 1 \%$. The 
power of each interference type relative to the noise power $P_\mathrm{int}$ is 
used for RRA since the true interference power is not known. 
For the network performance evaluation, the true interference 
power is calculated according to the distribution of all users in all cells.
All interferers can be either OSC or GMSK modulated, depending on the decisions made by the RRA algorithm.

The network simulator first generates new cells and then randomly distributes
users in the cells. On a cell per cell basis one of the RRA algorithms
from Section~\ref{sec:rra} is executed. The logical channels are then
randomly assigned to the physical channels. 
For all users in all cells, realizations of the small-scale fading are generated 
for each of the $4$ bursts involved in a frame,
and the FER is evaluated according to Section~\ref{sec:l2sysmap}.
Then, all users are removed from the cells and the procedure starts from the beginning
with the random distribution of new users in the cells,
i.e., there is no simulation over time.
The total number of users in all cells is fixed and the users
are randomly dropped into the total area which results in
an average number of users per cell $N_\mathrm{user}$.

\begin{figure}\begin{center}%
\subfigure[FER.\label{abb:vamos_no_vamos_fer}]{%
\resizebox{\subfigwidth}{!}{%
\psfragfig{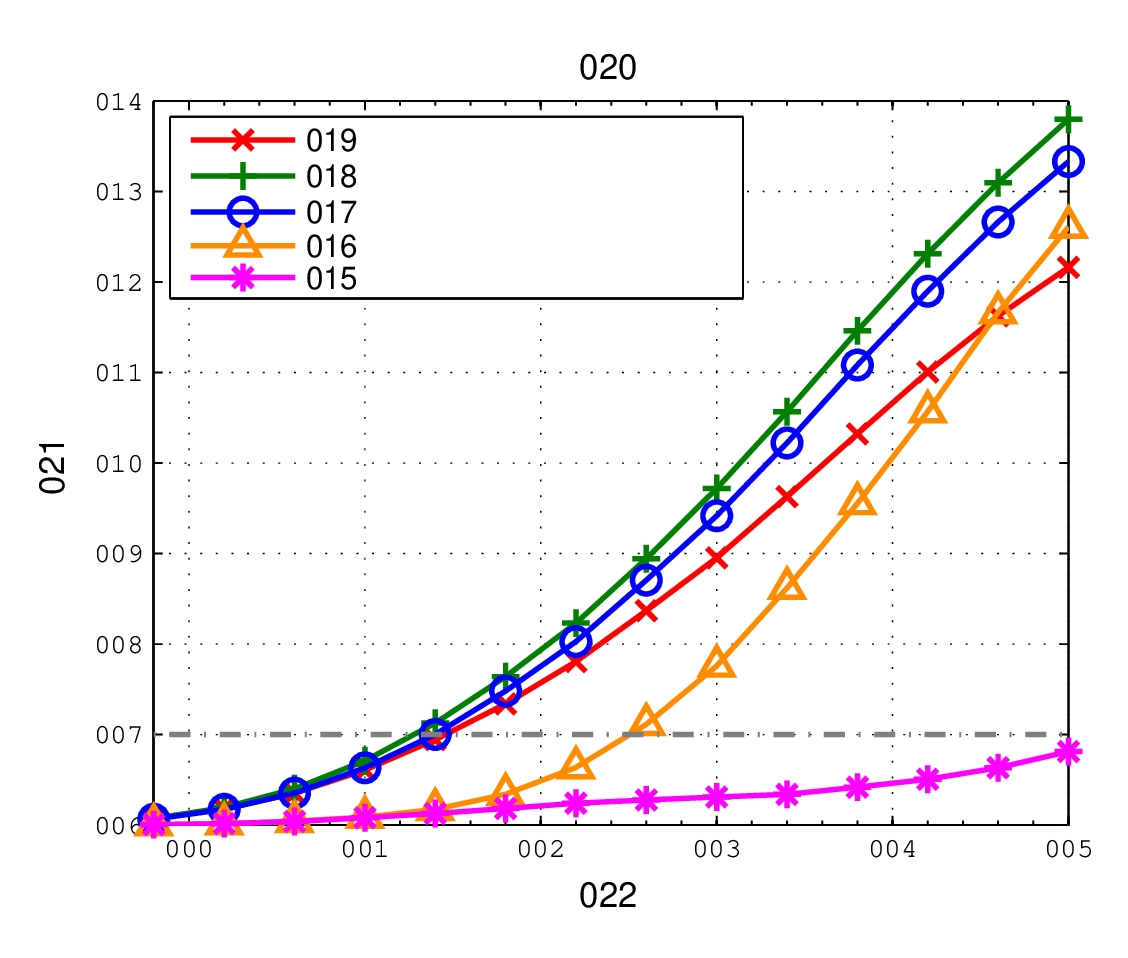}}
}%
\hspace{\subfigspace}
\subfigure[Average transmit power of BS per MS.\label{abb:vamos_no_vamos_power}]{%
\resizebox{\subfigwidth}{!}{%
\psfragfig{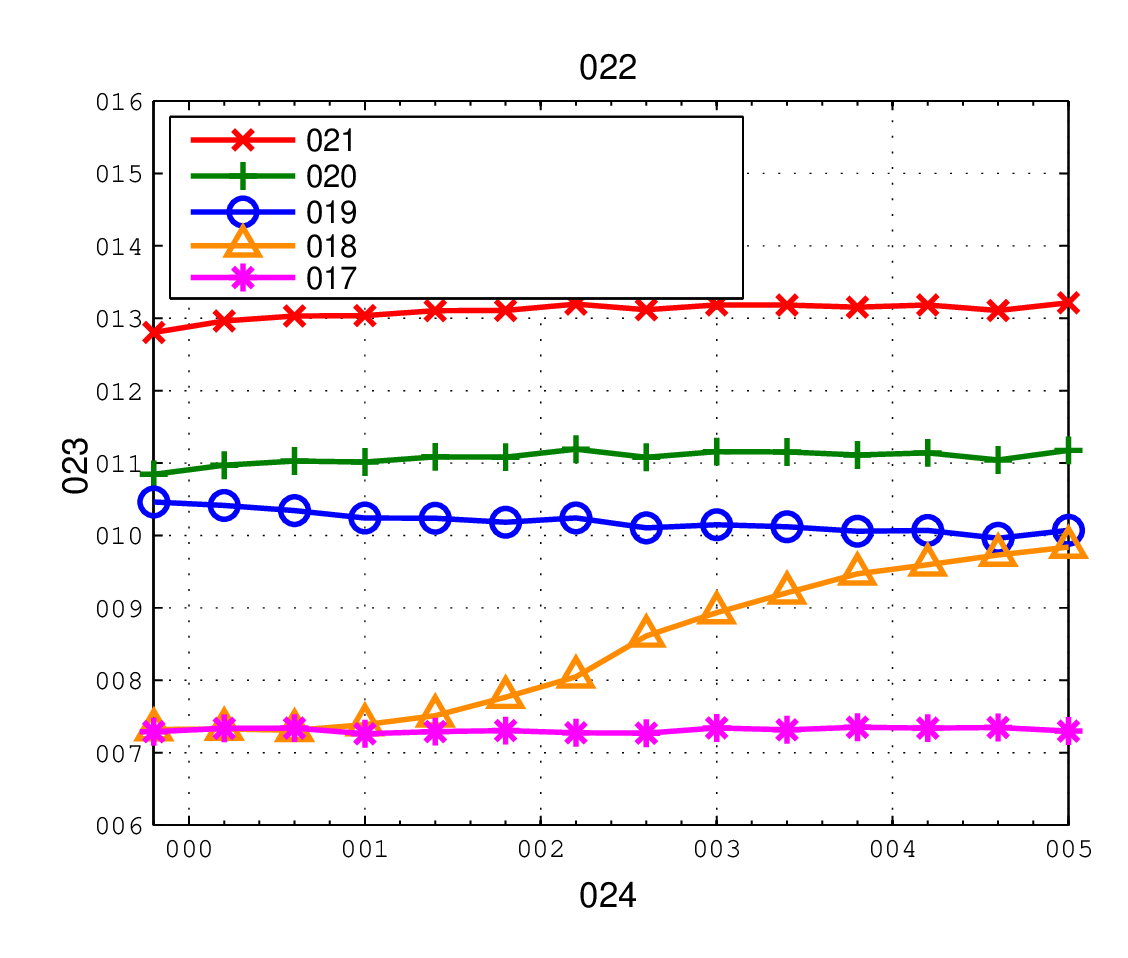}}
}%
\caption{FER and transmit power vs. average number of users per cell for VAMOS vs. no-VAMOS scenario, $P_\mathrm{int} = 10$~dB above noise power, MIC receiver.}%
\label{abb:vamos_no_vamos}
\vspace*{\afterfigspace}
\end{center}\end{figure}%
Figs.~\ref{abb:vamos_no_vamos_fer} and \ref{abb:vamos_no_vamos_power} depict the FER and the average transmit power 
of the BS per MS over the average
number of users per cell $N_\mathrm{user}$, respectively. 
A MIC receiver was used at all MSs. The lines marked with
``random pairing, SCPIR = 0 dB'' show the performance of random user pairing and 
equal power allocation within each pair ($\mathrm{SCPIR} = 0$~dB). 
Fig.~\ref{abb:vamos_no_vamos_power} shows that by 
optimizing the SCPIR for all random pairs (``SCPIR optim.''), a power
reduction of nearly $2$~dB is possible. The additional power saving enabled by
employing optimal user pairing (``user pairing'') compared to random user pairing and SCPIR 
optimization is about $0.5$~dB, and increases with the number of users available 
for user pairing. For these three cases all users were forced
to always transmit over OSC channels. One can see that the power required
for transmission is about $3$~dB higher than for transmission without VAMOS.
This is due to the fact that the power allocation for a single user only
needs to achieve the SNR target for that user, not for both users, cf. \eqref{eq:maxpower}.
When only pairing users that would be blocked if no OSC was used (``POOB''), one can see
that the necessary power lies between no-VAMOS and VAMOS with user pairing.
For a low average number of users in the cell, e.g. between $N_\mathrm{user} = 4$ and $8$, the necessary
power is equal to that of the no-VAMOS case. By increasing $N_\mathrm{user}$, the
number of users that receive their signal via OSC transmission increases and thereby also the
necessary transmit power. 

\begin{figure}\begin{center}%
\subfigure[FER.\label{abb:p_int_mod_fer}]{%
\resizebox{\subfigwidth}{!}{%
\psfragfig{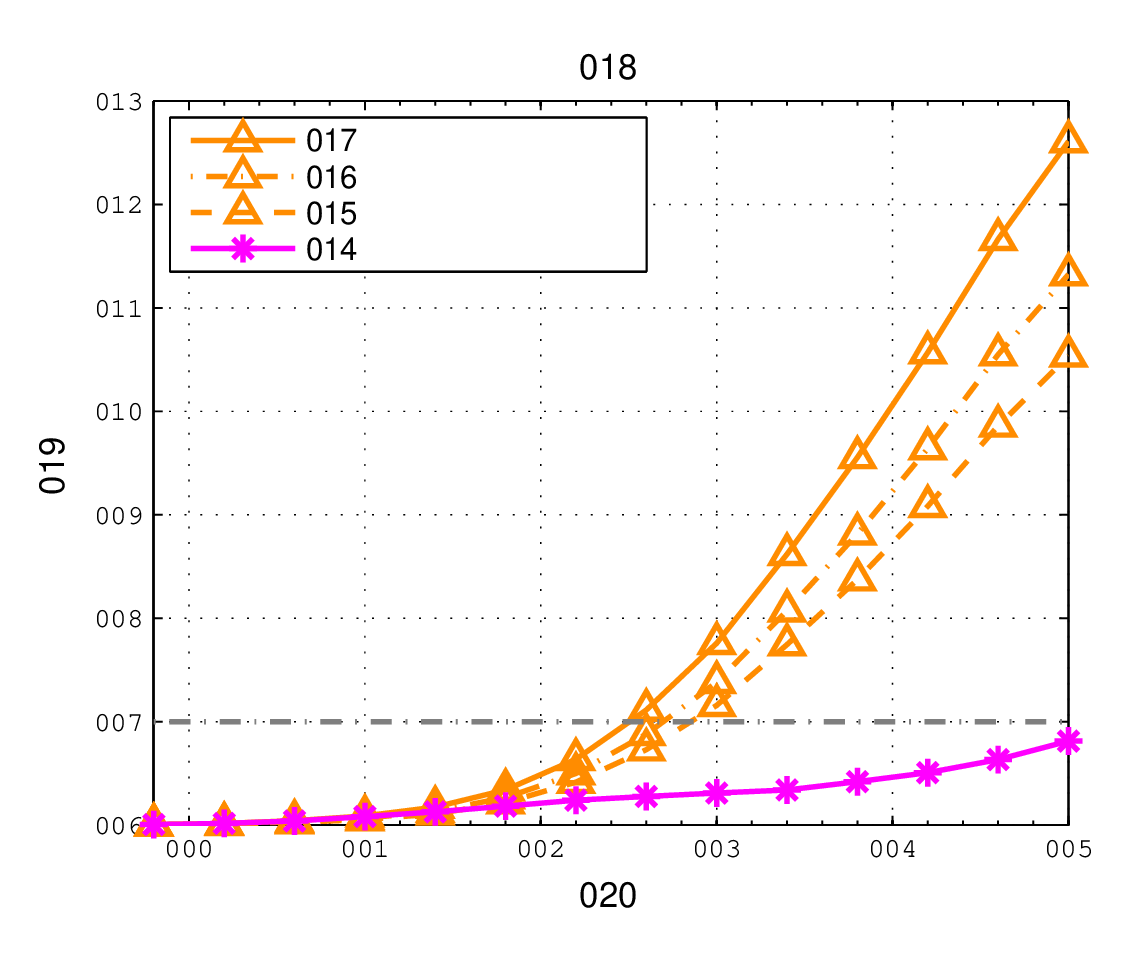}}
}%
\hspace{\subfigspace}
\subfigure[Average transmit power of BS per MS.\label{abb:p_int_mod_power}]{%
\resizebox{\subfigwidth}{!}{%
\psfragfig{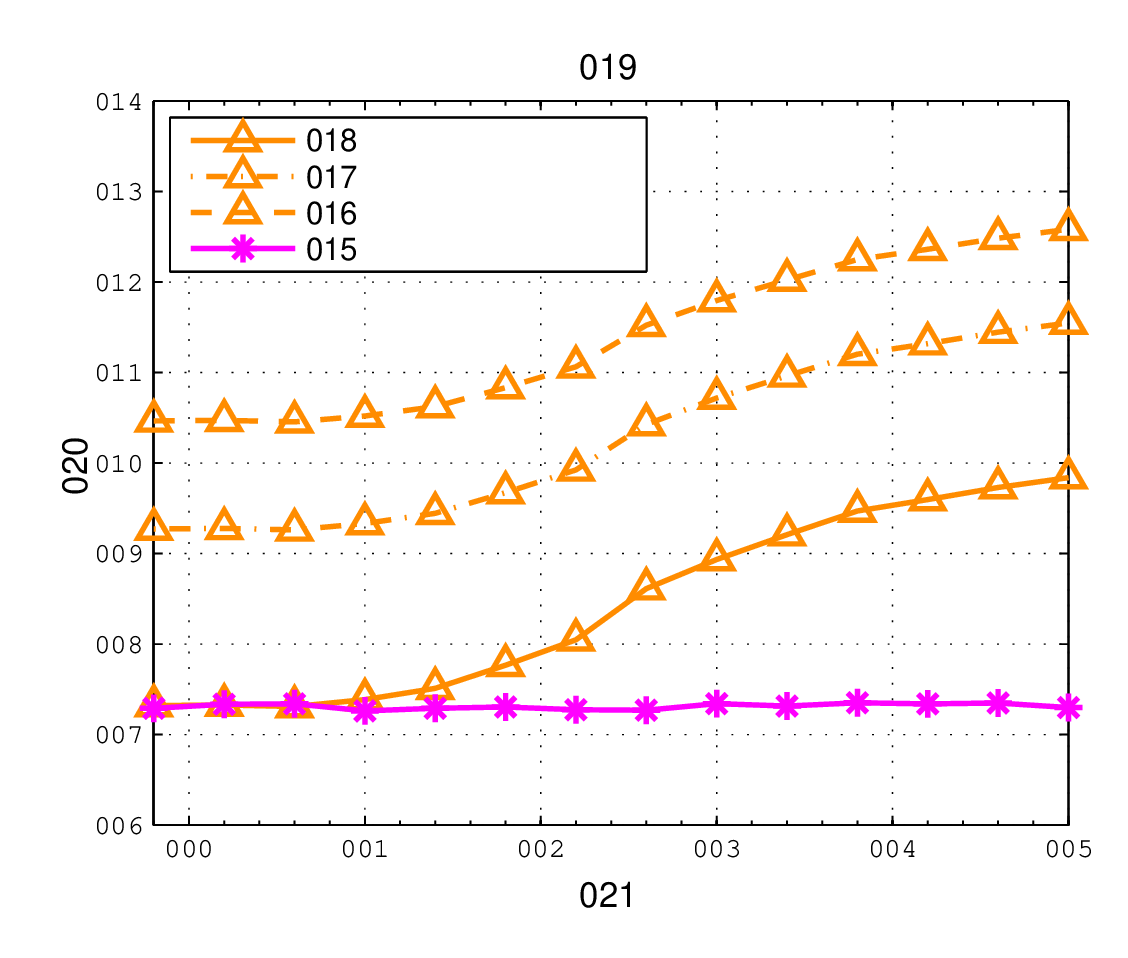}}
}%
\caption{FER and transmit power versus average number of users per cell for different $P_\mathrm{int}$ values, MIC receiver.}%
\vspace*{\afterfigspace}
\label{abb:p_int_mod}
\end{center}\end{figure}

The
FER of the different pairing strategies is depicted in 
Fig.~\ref{abb:vamos_no_vamos_fer}. One can see that an
$\mathrm{FER}_\mathrm{thr}$ (dash dotted line) of $1~\%$ cannot be fulfilled for a
high number of users in the cell and OSC transmission. When the load in the cells increases,
also the co-channel interference increases dramatically. The MIC receiver
that is employed for all cases cannot cancel all interferers anymore which
leads to an increased FER. Even for the no-VAMOS case, where also the MIC receiver is
used, the FER approaches the threshold for a high number of users. 
For the user pairing algorithms that try to
pair as many users as possible the interference is very often an OSC signal
that cannot be cancelled by the MIC receiver. This can also
be seen for the MTS-2 scenario investigated in Fig.~\ref{abb:MTS2_AFS59}. For the cases with high load,
the FER performance of user pairing is worse than without user pairing.
The resource allocation has been optimized for a fixed value 
of interference power $P_\mathrm{int}$. Different values for
$P_\mathrm{int}$ will be considered in Fig.~\ref{abb:p_int_mod}. However,
the actual interference power caused by the different algorithms
exceeds this value already for a medium system load. 
One can also observe that the lower transmit power that is achieved by 
user pairing compared to random pairing,  
results in a worse FER for high load.

\begin{figure}\begin{center}%
\resizebox{\figwidth}{!}{%
\psfragfig{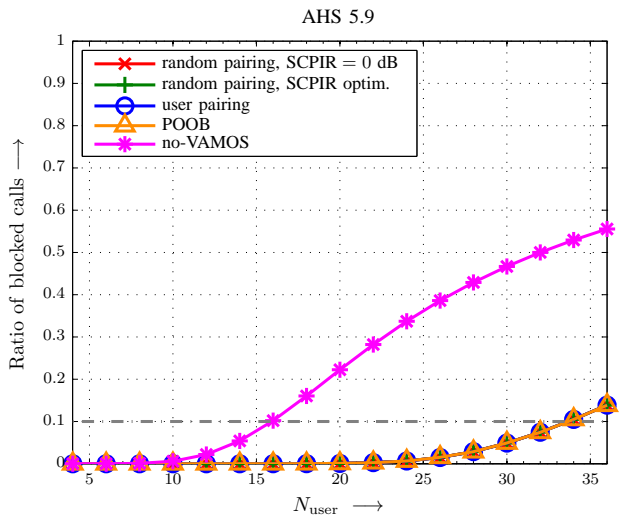}}
\caption{Ratio of blocked calls versus average number of users per cell for VAMOS vs. no-VAMOS scenario, $P_\mathrm{int} = 10$~dB above noise power, MIC receiver.}%
\label{abb:vamos_no_vamos_blocked_calls}
\vspace*{\afterfigspace}
\end{center}\end{figure}%
From Fig.~\ref{abb:vamos_no_vamos} one could come to the conclusion
that no-VAMOS would be the better choice. However, when taking into
account the number of blocked calls, depicted in 
Fig.~\ref{abb:vamos_no_vamos_blocked_calls}, the main benefit of 
VAMOS is revealed.
Here, we assume a call is blocked if not enough logical channels are 
available to schedule this call. For our simulations, $K = 8$ physical 
channels are available per cell. Compared to full rate, with the half rate codec 
the number of available logical channels per cell is doubled. As one can see from Fig.~\ref{abb:vamos_no_vamos_blocked_calls},
the percentage of blocked calls increases very fast for the no-VAMOS
case. Already for $N_\mathrm{user} = 16$ the percentage of blocked calls exceeds 
$10~\%$ (dash dotted line). In contrast, by employing the OSC concept, $10~\%$ blocked calls occur for 
$N_\mathrm{user} = 33.7$. The average number of users for a given percentage 
of blocked calls can be more than doubled
by doubling the number of available channels with VAMOS.
This can be explained with the Erlang B formula for the blocking probability,
which states that by increasing the number of available channels the blocking probability 
decreases for the same relative load.
This capacity gain is the reason why the OSC concept was introduced in GSM. 

A solution to overcome the undesirable FER behavior observed in Fig.~\ref{abb:vamos_no_vamos} is to
increase $P_\mathrm{int}$ for RRA, which will result in a higher power
consumption. Figs.~\ref{abb:p_int_mod_fer} and \ref{abb:p_int_mod_power} show the FER and transmit power,
respectively, for different values of $P_\mathrm{int}$. It can be
seen that by increasing the assumed interference power also the
transmit power is increased which has a positive influence on the
FER performance. Still, for a load higher than $19$ users, the FER threshold cannot be
satisfied anymore for a VAMOS transmission. 

\begin{figure}\begin{center}%
\subfigure[FER.\label{abb:no_dtx_fer}]{%
\resizebox{\subfigwidth}{!}{%
\psfragfig{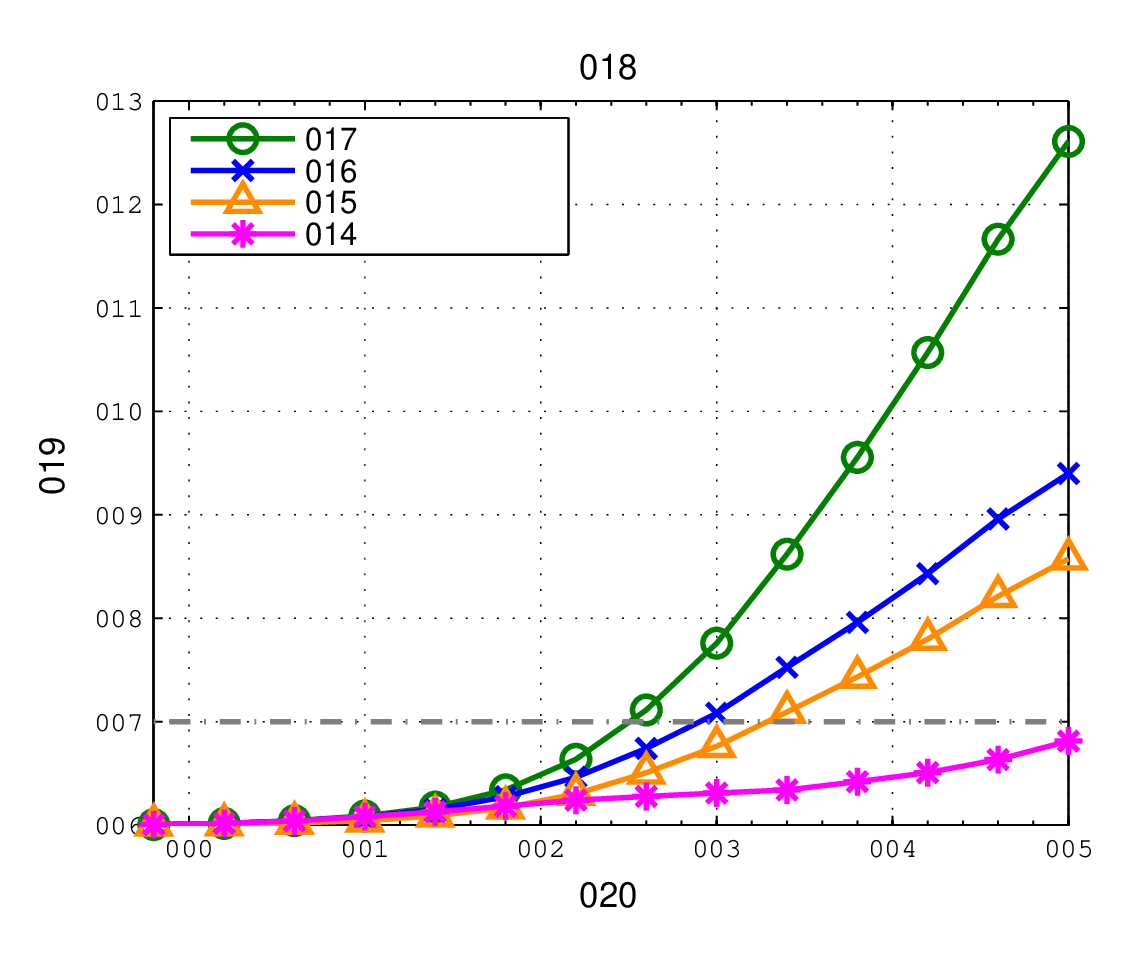}}
}%
\hspace{\subfigspace}
\subfigure[Average transmit power of BS per MS.\label{abb:no_dtx_power}]{%
\resizebox{\subfigwidth}{!}{%
\psfragfig{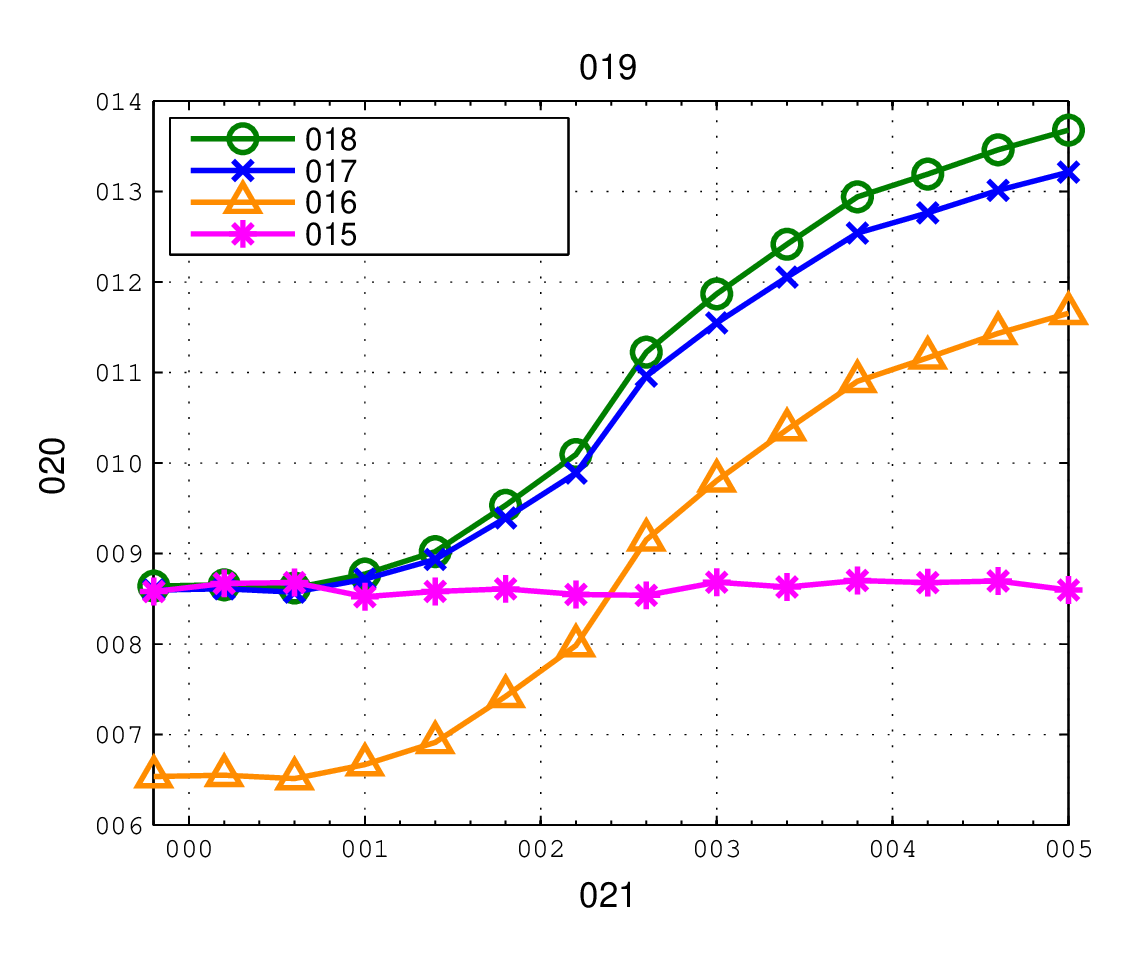}}
}%
\caption{FER and transmit power versus average number of users per cell for different receivers, $P_\mathrm{int} = 10$~dB above noise power, POOB user pairing.}%
\vspace*{\afterfigspace}
\label{abb:no_dtx}
\end{center}\end{figure}%
Therefore, the key to avoid the undesirable FER behavior without increasing
$P_\mathrm{int}$ is to use an enhanced receiver in the MSs. 
Figs.~\ref{abb:no_dtx_fer} and \ref{abb:no_dtx_power}
show the FER and transmit power, respectively, for POOB user pairing
if MIC, SIC, and V-MIC receivers
are employed at the MSs, respectively. 
Due to the better interference cancellation capabilities
of the SIC and V-MIC receivers, the FER 
is much lower than for the MIC receiver. For the SIC receiver 
a small transmit power saving compared to MIC can be achieved,
whereas for V-MIC the RRA can reduce the transmit power significantly.

\begin{figure}\begin{center}%
\subfigure[FER.\label{abb:dtx_fer}]{%
\resizebox{\subfigwidth}{!}{%
\psfragfig{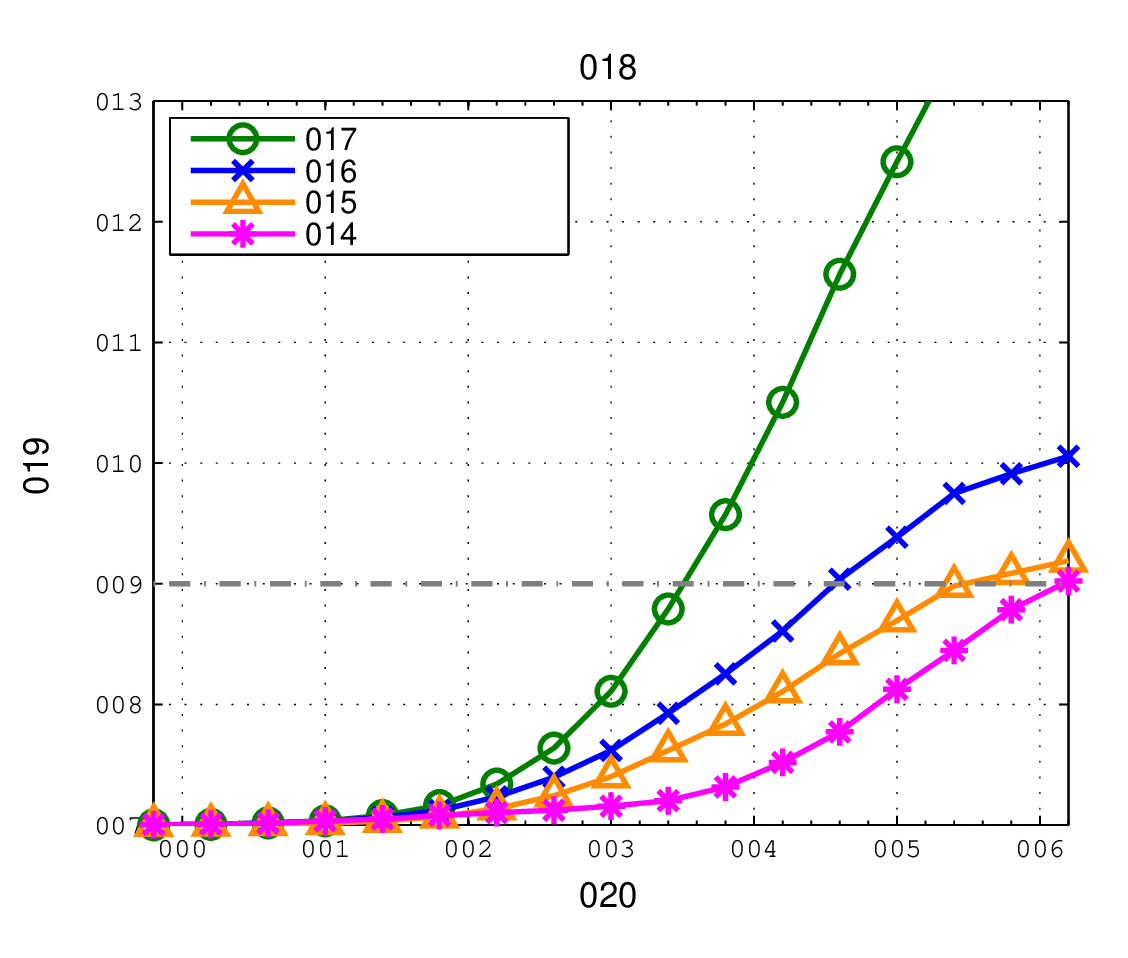}}
}%
\hspace{\subfigspace}
\subfigure[Average transmit power of BS per MS.\label{abb:dtx_power}]{%
\resizebox{\subfigwidth}{!}{%
\psfragfig{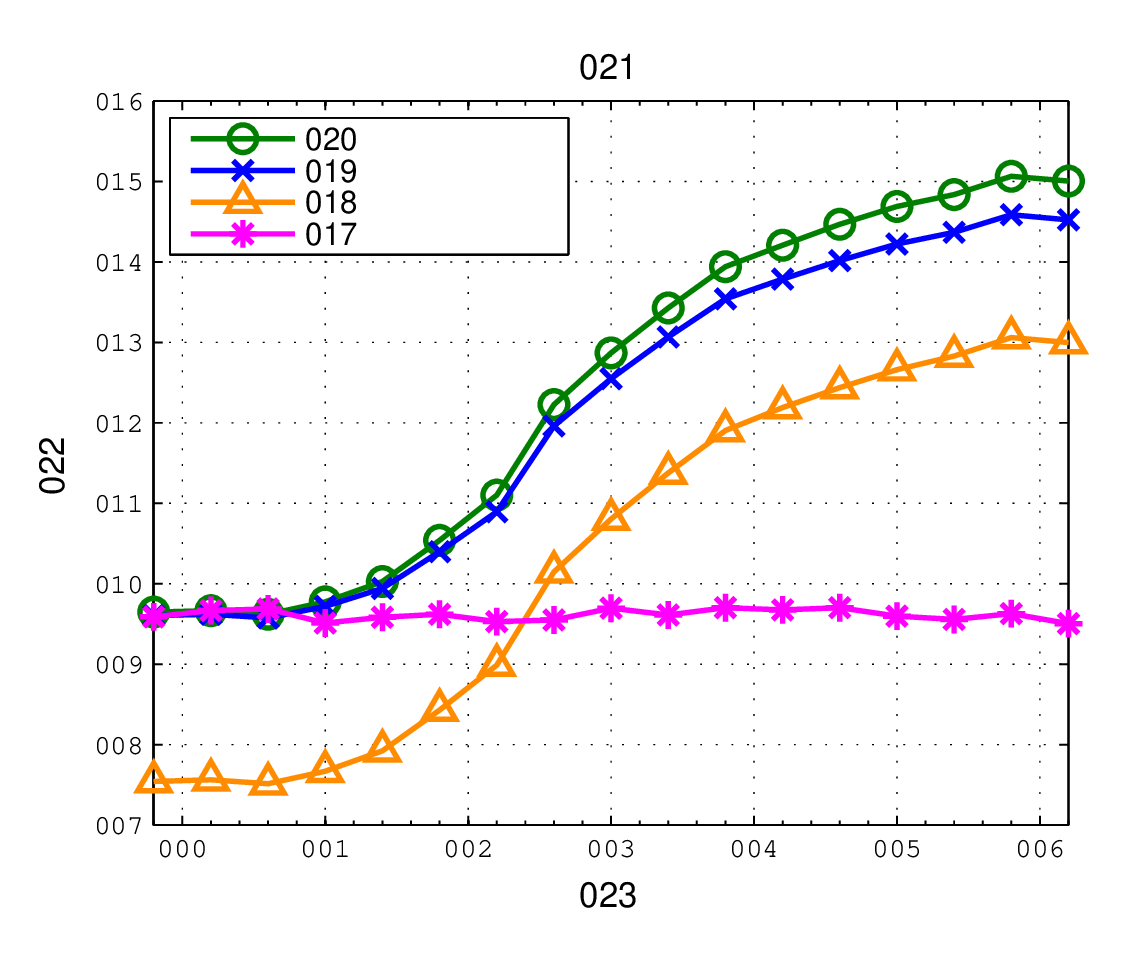}}
}%
\caption{FER and transmit power versus average number of users per cell for different VAMOS receivers, $P_\mathrm{int} = 10$~dB above noise power, DTX enabled, POOB user pairing.}%
\vspace*{\afterfigspace}
\label{abb:dtx}
\end{center}\end{figure}%
The FER and transmit power, respectively, for a more realistic scenario with enabled discontinuous transmission (DTX),
where a scheduled user does not transmit due to no speech 
activity with a probability of $40$\%, are depicted in
Figs.~\ref{abb:dtx_fer} and \ref{abb:dtx_power}. The POOB pairing strategy has been used for
the VAMOS results. The interference situation is now more relaxed
compared to the case without DTX. Within one pair, only with a
probability of $36$\% both users are active, while both users
of one pair are silent with a probability of $16$\%. This means
that strong interference by OSC users does not occur very often.
Furthermore, when the second user is not present, the receiver
can use its interference rejection capabilities to better suppress 
co-channel interferers from other cells.
With the V-MIC receiver it is now possible to keep the FER
below $1$\% for $N_\mathrm{user} \leq 32$. This enables a very high user load in the system.

\begin{figure}\begin{center}%
\subfigure[FER.\label{abb:hot_spot_FER}]{%
\resizebox{\subfigwidth}{!}{%
\psfragfig{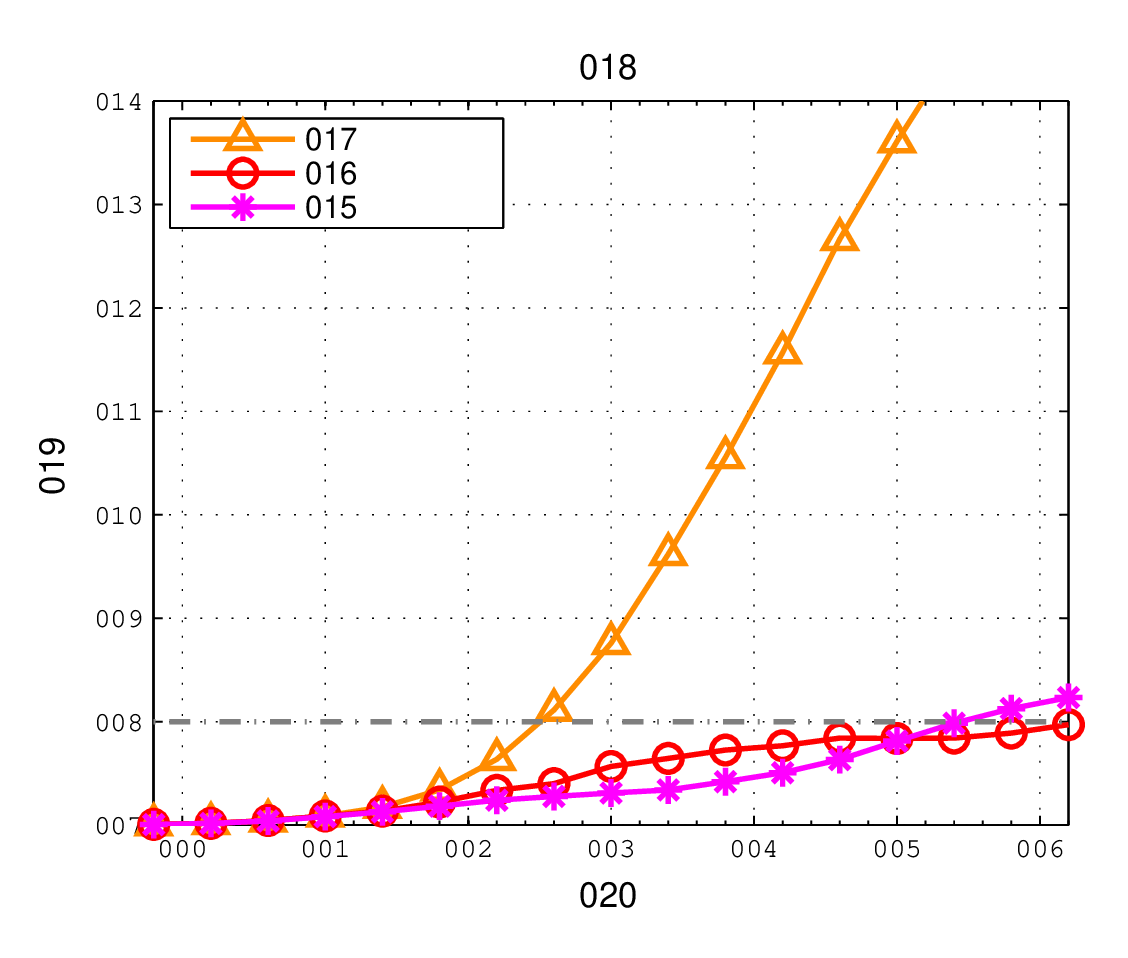}}
}%
\hspace{\subfigspace}
\subfigure[Average transmit power of BS per MS.\label{abb:hot_spot_power}]{%
\resizebox{\subfigwidth}{!}{%
\psfragfig{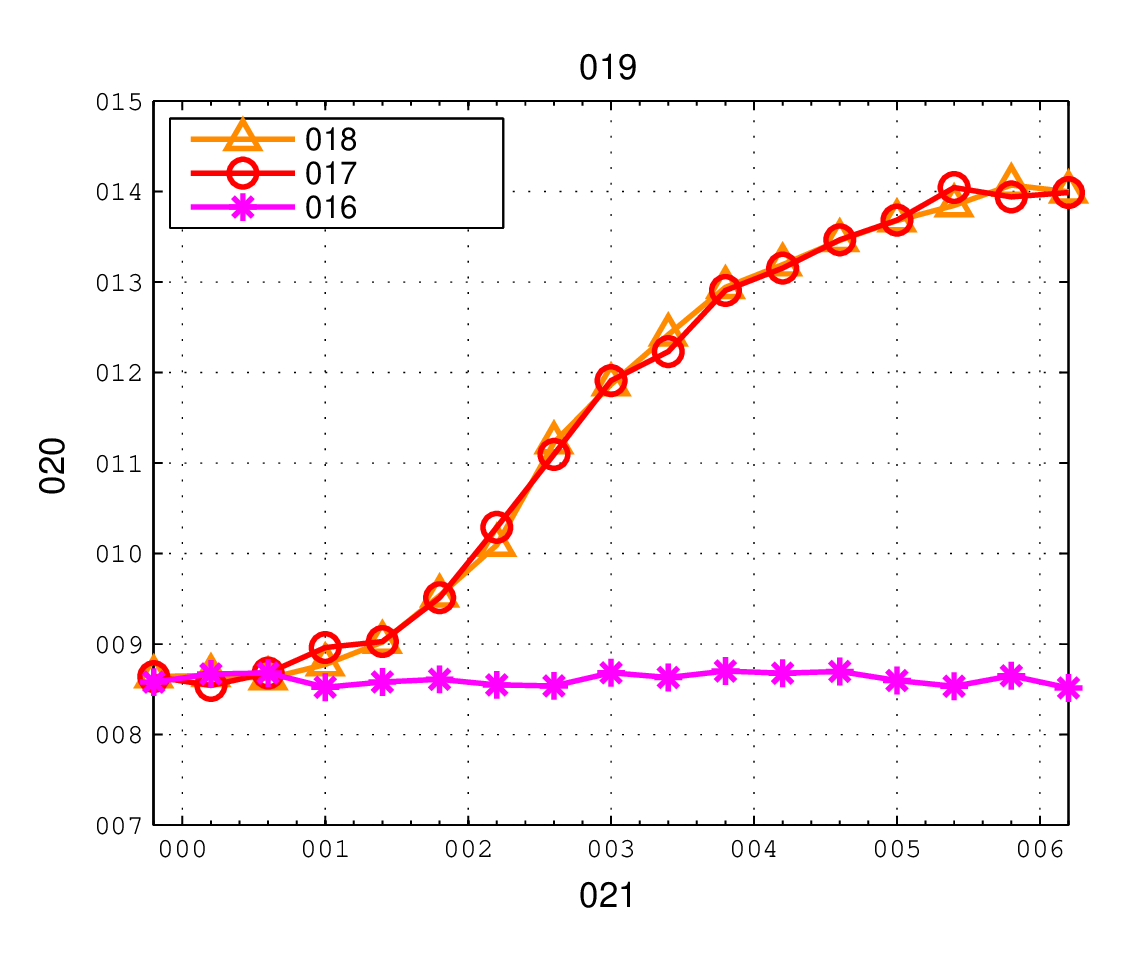}}
}%
\caption{FER and transmit power versus average number of users per cell, MIC receiver, hot spot scenario, $P_\mathrm{int} = 10$~dB above noise power.}%
\vspace*{\afterfigspace}
\label{abb:hot_spot}
\end{center}\end{figure}%
In Figs.~\ref{abb:hot_spot_FER} and \ref{abb:hot_spot_power}, the FER and transmit power, respectively, for a hot spot scenario are depicted.
A comparison is made with a no-VAMOS GMSK modulated transmission and 
OSC transmission with POOB user pairing in all cells. DTX is deactivated in all cases
and a MIC receiver is employed in all MSs. 
For the hot spot scenario the cell layout is not changed compared to the scenarios in Figs.~\ref{abb:vamos_no_vamos}-\ref{abb:dtx}.
However, for the hot spot scenario only one cell (the hot spot)
uses OSC transmission and all other co-channel cells use
legacy GSM with GMSK modulation. Only the FER of the
hot spot cell is shown here.
One can also view this scenario as perfect frequency assignment, where
all co-channel cells schedule the OSC users in such a way that only
GMSK interferers are present for an OSC pair. 
It can be observed that 
the resulting FER for the hot spot scenario is always below $1$\%. 
This shows that the FER reduction
due to the missing OSC interference is significant.
However, this also suggests that in other work, such as \cite{Molteni2011},
where only GMSK interference is assumed, the actually achievable performance
of a fully loaded VAMOS network may be overestimated.
We note that especially for the case of a high load in all cells, perfect 
frequency assignment over all cells guaranteeing only GMSK interference 
for the OSC users is impossible.
To have a fair performance comparison of the hot spot scenario 
with POOB, we use for the hot spot scenario the same transmit
power allocation 
that is also used in the case of OSC interference for all cells. 
For $N_\mathrm{user} > 30$, the $\mathrm{FER}$ of the hot spot is even lower
than that of the no-VAMOS scenario with GMSK modulation. This is a consequence of the 
higher transmit power allocation in the hot spot scenario compared to the no-VAMOS
scenario.

\begin{table}%
\centering
\caption{Overall network capacity gain of OSC compared to non-OSC transmission with $P_\mathrm{int} = 10~\mathrm{dB}$.}
\begin{tabular}{lr}
\toprule  
         Scenario & OSC gain\\ 
         \midrule
POOB, MIC, no DTX &  $9.4$\%\\
POOB, SIC, no DTX   & $21.9$\%\\
POOB, V-MIC, no DTX & $34.4$\%\\
POOB, MIC, DTX &  $40.6$\%\\
POOB, SIC, DTX  & $75.0$\%\\
POOB, V-MIC, DTX & $103$\% \\ 
hot spot, MIC, no DTX & $112$\%\\
\bottomrule 
\end{tabular}
\label{tab:cap_gain} 
\end{table}%
Table~\ref{tab:cap_gain} summarizes the overall network capacity gain of
OSC transmission for different parameters compared to no-VAMOS transmission with 
the MIC receiver, where $P_\mathrm{int} = 10$~dB is valid for all cases.
The gain is computed by comparing the maximum number of users with 
$\mathrm{FER} < 1$\% and  $\mathrm{blocked~calls} < 10$\% for each case. As a reference we use
the no-VAMOS transmission, where $10$\% blocked calls occur for $N_\mathrm{user} = 16$.
In all considered cases, a network capacity gain of OSC compared to no-VAMOS can be observed. 
With DTX enabled, the new V-MIC exhibits a network capacity gain of
more than $100$\%. For the hot spot with a MIC receiver, when OSC is only used in one cell,
even more than $100$\% network capacity
gain can be achieved.

\section{Conclusions}
\label{sec:conclusion}

In this paper, different receiver concepts for 
OSC downlink transmission used in VAMOS have
been introduced. Channel estimation
algorithms have been proposed, and
different receivers for OSC transmission
over frequency-selective fading channels with interference have
been introduced and compared with respect to their frame error rate performance.
It was shown that the novel V-MIC receiver exhibits a significant performance
improvement compared to receivers from the literature.
A practical transmit power allocation and user pairing algorithm
has been proposed as well. Different receiver concepts
have been evaluated in a network scenario using
the proposed radio resource allocation algorithm.
The benefits of discontinuous transmission and
the strong dependence of
the VAMOS downlink performance 
on the type of interference have been shown. Capacity gains of more than
$100$\% compared to no-VAMOS transmission can be obtained with the novel V-MIC receiver
in a realistic network scenario.

In summary, powerful solutions for the reception of OSC downlink signals
have been developed in this paper. These receivers combined with the proposed radio resource allocation
schemes achieve a very good performance even if exact knowledge of
the interference situation is not available for resource allocation.%
\ifCLASSOPTIONonecolumn%
\else%
\balance
\fi%


\bibliographystyle{IEEEtran}
\bibliography{IEEEabrv,./muros}

\end{document}

%% file: macros.tex
\newcommand{\vv}[1]{\boldsymbol{\mathrm{#1}}}
\newcommand{\mm}[1]{\boldsymbol{\mathrm{#1}}}

\newcommand{\expect}{\mathcal{E}}

\newcommand{\dsig}[2]{ #1 [ #2 ] }              

\newcommand{\dmatilde}[2]{\tilde{\mbox{\boldmath $ #1 $}} [ #2] }

\newcommand{\ma}[1]{\mathrm{\mathbf{#1}}} 

\newcommand{\argmin}[2]{\! \!
    \begin{array}[t]{c}
    \displaystyle {\rm argmin} \\[#1]
    \scriptstyle #2
    \end{array} \!
    }

\newcommand{\argmax}[2]{\! \!
    \begin{array}[t]{c}
    \displaystyle {\rm argmax} \\[#1]
    \scriptstyle #2
    \end{array} \!
    }

\makeatletter
\long\def\unmarkedfootnote#1{{\long\def\@makefntext##1{##1}\footnotetext{#1}}}
\makeatother


%% file: equalizer_tikz.tex
\tikzstyle{int}=[draw, fill=white, inner sep=0.4cm, text centered, drop shadow]
\tikzstyle{kreis}=[circle, draw, fill=black, inner sep=1.5pt, node distance=2.3cm]
\tikzstyle{summe}=[circle, inner sep=1pt, draw, fill=white, text centered]
\tikzstyle{init} = [pin edge={to-,thin,black}]
\tikzstyle{boller} = [circle, draw, fill=white, inner sep=1.5pt]
\tikzstyle{boller_black} = [circle, draw, fill=black, inner sep=1pt]

\begin{tikzpicture}[node distance=2.5cm,auto,>=latex']
	\node [boller, node distance=0cm] (anfang){} ;
	\node [boller_black, right of = anfang, node distance=0.75cm](abzweig){};
	\node [above of=anfang, node distance=.35cm]{$r_o[k]$};
	\node [int, right of=anfang, node distance=2.5cm] (f1) {$F_o(z)$};
  \node [int, right of=f1](P) {$\mathcal{P}_c\{\cdot\}$};	
	\node [summe, right of=P] (summe) {$+$};
	\node [boller, right of=summe, node distance=2.5cm] (ende){} ;
	\node [above of=ende, node distance=.35cm]{$y_o[k] = a_o[k-k_0]+e_o[k]$};
	\node [int, below right of=summe, node distance=2.9cm] (B11) {$B_{o,o}(z)$};
	\node [int, below of=B11, node distance = 1.7cm] (B12) {$B_{o,p}(z)$};
	\node [boller, right of=B11, node distance = 2cm](a1){};
	\node [above of=a1, node distance=.35cm]{$a_o[k]$};	
	\node [boller, right of=B12, node distance = 2cm](a2){};
	\node [above of=a2, node distance=.35cm]{$a_p[k]$};

    \path[->] (anfang) edge node {} (f1);
    \path[->] (f1) edge node {} (P);
    \path[->] (P) edge node {$u_o[k]$} (summe);
    \path[-] (summe) edge node {} (ende);
    \path[->] (a1) edge node {} (B11);
    \path[->] (a2) edge node {} (B12);
		\draw[->] (B11.west) -- ++(-0.3cm,0cm) -- node[at end,below right] {$-$} (summe);
		\draw[->] (B12) -| node[at end,below left] {$-$} (summe);

\node [int, below of=f1, node distance=5.5cm] (f2) {$F_p(z)$};
  \node [int, right of=f2](P) {$\mathcal{P}_c\{\cdot\}$};	
	\node [summe, right of=P] (summe) {$+$};
	\node [boller, right of=summe, node distance=2.5cm] (ende){} ;
	\node [above of=ende, node distance=.35cm]{$y_p[k] = a_p[k-k_0]+e_p[k]$};
	\node [int, below right of=summe, node distance=2.9cm] (B21) {$B_{p,o}(z)$};
	\node [int, below of=B21, node distance = 1.7cm] (B22) {$B_{p,p}(z)$};
	\node [boller, right of=B21, node distance = 2cm](a1){};
	\node [above of=a1, node distance=.35cm]{$a_o[k]$};	
	\node [boller, right of=B22, node distance = 2cm](a2){};
	\node [above of=a2, node distance=.35cm]{$a_p[k]$};

		\draw[->] (abzweig.south) |- node[at end,below left] {} (f2.west);
    \path[->] (f2) edge node {} (P);
    \path[->] (P) edge node {$u_p[k]$} (summe);
    \path[-] (summe) edge node {} (ende);
    \path[->] (a1) edge node {} (B21);
    \path[->] (a2) edge node {} (B22);
		\draw[->] (B21.west) -- ++(-0.3cm,0cm) -- node[at end,below right] {$-$} (summe);
		\draw[->] (B22) -| node[at end,below left] {$-$} (summe);

%
%
%
%
%
%
%
%
%
%
%
%
%

\end{tikzpicture}

%% file: vamos_transactions.bbl
\begin{thebibliography}{10}
\providecommand{\url}[1]{#1}
\csname url@samestyle\endcsname
\providecommand{\newblock}{\relax}
\providecommand{\bibinfo}[2]{#2}
\providecommand{\BIBentrySTDinterwordspacing}{\spaceskip=0pt\relax}
\providecommand{\BIBentryALTinterwordstretchfactor}{4}
\providecommand{\BIBentryALTinterwordspacing}{\spaceskip=\fontdimen2\font plus
\BIBentryALTinterwordstretchfactor\fontdimen3\font minus
  \fontdimen4\font\relax}
\providecommand{\BIBforeignlanguage}[2]{{%
\expandafter\ifx\csname l@#1\endcsname\relax
\typeout{** WARNING: IEEEtran.bst: No hyphenation pattern has been}%
\typeout{** loaded for the language `#1'. Using the pattern for}%
\typeout{** the default language instead.}%
\else
\language=\csname l@#1\endcsname
\fi
#2}}
\providecommand{\BIBdecl}{\relax}
\BIBdecl

\bibitem{Gerstacker05}
R.~Meyer, W.~Gerstacker, R.~Schober, and J.~B. Huber, ``{A} {S}ingle {A}ntenna
  {I}nterference {C}ancellation {A}lgorithm for {I}ncreased {GSM} {C}apacity,''
  \emph{IEEE Transactions on Wireless Communications}, vol.~5, no.~7, pp.
  1616--1621, 2006.

\bibitem{Chevalier2006}
P.~Chevalier and F.~Pipon, ``{N}ew {I}nsights {I}nto {O}ptimal {W}idely
  {L}inear {A}rray {R}eceivers for the {D}emodulation of {BPSK}, {MSK}, and
  {GMSK} {S}ignals {C}orrupted by {N}oncircular {I}nterferences - {A}pplication
  to {SAIC},'' \emph{IEEE Transactions on Signal Processing}, vol.~54, pp.
  870--883, 2006.

\bibitem{nokia:07}
Nokia, \emph{3{GPP} {T}doc {GP}-071792, {V}oice {C}apacity {E}volution with
  {O}rthogonal {S}ub {C}hannels}, 3GPP TSG GERAN \#36, Vancouver, Canada, Nov.
  2007.

\bibitem{chen:09}
X.~Chen, Z.~Fei, J.~Kuang, L.~Liu, and G.~Yang, ``{A} {S}cheme of {M}ulti-user
  {R}eusing {O}ne {S}lot on {E}nhancing {C}apacity of {GSM}/{EDGE}
  {N}etworks,'' in \emph{Proc. of 11th IEEE Singapore International Conference
  on Communication Systems (ICCS 2008)}, Singapore, Nov. 2008, pp. 1574--1578.

\bibitem{OSC_chap2011}
M.~Säily, J.~Hulkkonen, K.~Pedersen, C.~Juncker, R.~Paiva, R.~Iida,
  O.~Piirainen, S.~Sundaralingam, A.~Loureiro, J.~Helt-Hansen, R.~Domingos, and
  F.~Tavares, \emph{{GSM}/{EDGE}: {E}volution and {P}erformance}.\hskip 1em
  plus 0.5em minus 0.4em\relax John Wiley \& Sons, 2011, ch. Orthogonal
  Sub-Channels with AMR/DARP, pp. 235 -- 276.

\bibitem{Paiva2012}
R.~C.~D. Paiva, R.~D. Vieira, R.~Iida, F.~M. Tavares, M.~Saily, J.~Hulkkonen,
  R.~Jarvela, and K.~Niemela, ``{GSM} {V}oice {E}volution {U}sing {O}rthogonal
  {S}ubchannels,'' \emph{IEEE Communications Magazine}, vol.~50, no.~12, pp.
  80--86, 2012.

\bibitem{TR45914:940}
\emph{3{GPP} {TR} 45.914, {C}ircuit switched voice capacity evolution for
  {GSM}/{EDGE} {R}adio {A}ccess {N}etwork ({GERAN}), version 9.4.0}, Nov. 2010.

\bibitem{Ruder2011}
M.~A. Ruder, R.~Schober, and W.~H. Gerstacker, ``{C}ramer-{R}ao {L}ower {B}ound
  for {C}hannel {E}stimation in a {MUROS}/{VAMOS} {D}ownlink {T}ransmission,''
  in \emph{Proc. IEEE 22nd Int. Personal Indoor and Mobile Radio Communications
  (PIMRC) Symp.}, 2011, pp. 1433--1437.

\bibitem{Vutukuri2011}
M.~G. Vutukuri, R.~Malladi, K.~Kuchi, and R.~D. Koilpillai, ``{SAIC} {R}eceiver
  {A}lgorithms for {VAMOS} {D}ownlink {T}ransmission,'' in \emph{Proc. 8th Int.
  Wireless Communication Systems (ISWCS) Symp.}, 2011, pp. 31--35.

\bibitem{Molteni2011}
D.~Molteni, M.~Nicoli, and M.~S{\"a}ily, ``{R}esource {A}llocation {A}lgorithm
  for {GSM}-{OSC} {C}ellular {S}ystems,'' in \emph{Proc. IEEE Int. Conf. on
  Communications (ICC)}, 2011, pp. 1--6.

\bibitem{Ruder2012}
M.~A. Ruder, R.~Meyer, H.~Kalveram, and W.~H. Gerstacker, ``{R}adio {R}esource
  {A}llocation for {OSC} {D}ownlink {C}hannels,'' in \emph{Proc. of 1st IEEE
  Int. Conf. on Communications in China (ICCC), Workshop on Smart and Green
  Communications \& Networks (SGCNet)}, 2012, pp. 113--118.

\bibitem{Meyer2009}
R.~Meyer, W.~H. Gerstacker, F.~Obernosterer, M.~A. Ruder, and R.~Schober,
  ``{E}fficient {R}eceivers for {GSM} {MUROS} {D}ownlink {T}ransmission,'' in
  \emph{Proc. IEEE 20th Int. Personal, Indoor and Mobile Radio Communications
  Symp. (PIMRC)}, 2009, pp. 2399--2403.

\bibitem{TS45004:1100}
\emph{3{GPP} {TS} 45.004, {M}odulation, version 11.0.0}, Sep. 2012.

\bibitem{crozier:91}
S.~Crozier, D.~Falconer, and S.~Mahmoud, ``{L}east {S}um of {S}quared {E}rrors
  ({LSSE}) {C}hannel {E}stimation,'' \emph{IEE Proceedings F}, vol. 138, pp.
  371--378, Aug. 1991.

\bibitem{brent:73}
R.~P. Brent, \emph{{A}lgorithms for {M}inimization without
  {D}erivatives}.\hskip 1em plus 0.5em minus 0.4em\relax Englewood Cliffs, New
  Jersey: Prentice-Hall, 1973.

\bibitem{fossorier:98}
M.~Fossorier, F.~Burkert, S.~Lin, and J.~Hagenauer, ``{O}n the {E}quivalence
  between {SOVA} and {M}ax-{L}og-{MAP} {D}ecodings,'' \emph{IEEE Communications
  Letters}, vol.~2, no.~5, pp. 137--139, May 1998.

\bibitem{koch:89}
W.~Koch and A.~Baier, ``{C}ombined {D}esign of {E}qualizer and {C}hannel
  {D}ecoder for {D}igital {M}obile {R}adio {R}eceivers,'' \emph{Proceedings of
  ITG Conference ,,Stochastische Modelle und Methoden in der
  Informationstechnik''}, pp. 263--270, Apr. 1989.

\bibitem{TS45005:9130}
\emph{3{GPP} {TS} 45.005, {R}adio {T}ransmission and {R}eception ({R}elease 9),
  version 9.13.0}, Mar. 2013.

\bibitem{saic_pat_pct}
R.~Meyer, W.~Gerstacker, and R.~Schober, ``{M}ethod for {C}ancelling
  {I}nterference during {TDMA} {T}ransmission and/or {FDMA} {T}ransmission,''
  Patent, Dec., 2000.

\bibitem{SAIC2005VTC}
R.~Meyer, W.~Gerstacker, R.~Schober, and J.~B. Huber, ``{A} {S}ingle {A}ntenna
  {I}nterference {C}ancellation {A}lgorithm for {GSM},'' in \emph{Proc. of
  Vehicular Technology Conf. (VTC 2005-Spring)}, Stockholm, Sweden, May / June
  2005.

\bibitem{comres2010}
Com-Research, \emph{3{GPP} {T}doc {GP}-100083, {VAMOS} {D}ownlink {R}eceiver
  {P}erformance for {CCI} and {ACI} {S}cenarios}, 3GPP TSG GERAN \#45, Berlin,
  Germany, Mar. 2010.

\bibitem{Molteni2009}
D.~Molteni and M.~Nicoli, ``{A} {N}ovel {U}plink {R}eceiver for {GSM}/{EDGE}
  {S}ystems with {O}rthogonal {S}ub {C}hannel {F}eature,'' in \emph{Proc. of
  the Forty-Third Asilomar Conf. on Signals, Systems, and Computers}, 2009, pp.
  977--981.

\bibitem{zhang2005}
J.~Zhang, A.~Sayeed, and B.~Van~Veen, ``{R}educed-state {MIMO} {S}equence
  {D}etection with {A}pplication to {EDGE} {S}ystems,'' \emph{IEEE Transactions
  on Wireless Communications}, vol.~4, no.~3, pp. 1040--1049, May 2005.

\bibitem{Brueck2004274}
S.~Brueck, H.-J. Kettschau, and F.~Obernosterer,
  ``\BIBforeignlanguage{English}{{E}mission {R}eduction and {C}apacity
  {I}ncrease in {GSM} {N}etworks by {S}ingle {A}ntenna {I}nterference
  {C}ancellation},'' \emph{\BIBforeignlanguage{English}{AEU - International
  Journal of Electronics and Communications}}, vol.~58, no.~4, pp. 274--283,
  2004.

\bibitem{papadimitriou1998combinatorial}
C.~Papadimitriou and K.~Steiglitz, \emph{{C}ombinatorial {O}ptimization:
  {A}lgorithms and {C}omplexity}.\hskip 1em plus 0.5em minus 0.4em\relax Dover
  Publications, 1998.

\bibitem{DeCarvalho1997}
E.~De~Carvalho and D.~T.~M. Slock, ``{C}ramer-{R}ao {B}ounds for {S}emi-blind,
  {B}lind and {T}raining {S}equence based {C}hannel {E}stimation,'' in
  \emph{Proc. First IEEE Signal Processing Workshop Signal Processing Advances
  in Wireless Communications}, 1997, pp. 129--132.

\bibitem{Paiva2010}
R.~C.~D. Paiva, R.~Vieira, R.~Järvelä, R.~Iida, F.~Tavares, and M.~Säily,
  ``{I}mproving the {S}peech {Q}uality with {OSC}: {D}ouble {F}ull-{R}ate
  {P}erformance {A}ssessment,'' in \emph{Proc. IEEE 72nd Vehicular Technology
  Conf. Fall (VTC 2010-Fall)}, 2010, pp. 1--5.

\end{thebibliography}
